# CHERT: Cached Hamiltonian-Enhanced Ray Tracing in Viscoelastic Anisotropic Media

Muhammad F. T. Darwis, Department of Earth Sciences, Khalifa University of Science and Technology, Abu Dhabi, United Arab Emirates. E-mail: 100060597@ku.ac.ae ; ORCID: 0009-0009-8331-6572.

Fateh Bouchaala, Department of Earth Sciences, Khalifa University of Science and Technology, Abu Dhabi, United Arab Emirates. E-mail: fateh.bouchaala@ku.ac.ae; ORCID: 0000-0003-2368-3302.

Umair Bin Waheed, Geosciences Department, King Fahd University of Petroleum and Minerals, Dhahran, Saudi Arabia. E-mail: umair.waheed@kfupm.edu.sa

Mohamed Kamel Riahi, Department of Mathematics, Khalifa University of Science and Technology, Abu Dhabi, United Arab Emirates. E-mail: mohamed.riahi@ku.ac.ae

Corresponding author: Fateh Bouchaala, fateh.bouchaala@ku.ac.ae

Keywords: anisotropy, ray tracing, viscoelastic, traveltime, TTI.

Three sentence novelty statement: CHERT decouples the g*-Hamiltonian ray quantity computation from field-scale shortest-path ray tracing through reusable directional caches. The cache is indexed by wave mode, class, and sampled ray direction, and provides ray velocity, attenuation, and direction during trajectory updates without repetitive Hamiltonian evaluations. Controlled comparisons against MSPM demonstrate qP and qSH raypaths and traveltime fields consistency, qSV path-selection sensitivity, and significant computational time reduction, while the angular-coverage and nodewise benchmark test validates the CHERT's computational performance.

Running head: Cached Hamiltonian-Enhanced Ray Tracing.

ABSTRACT

Ray tracing in viscoelastic anisotropic media remains challenging because reliable energy velocities and ray quantities must be computed at the point scale and efficiently incorporated into whole-model traveltime and raypath calculations. Recent g*-Hamiltonian formulations have improved the qSV ray quantities and wavefront in viscoelastic anisotropic media. However, repeatedly evaluating the local Hamiltonian along candidate paths remains computationally expensive when modeling traveltime fields and raypaths. This study introduces a cached-Hamiltonian ray tracing (CHERT) method that decouples Hamiltonian node-scale evaluation from the shortest-path framework via precomputation and attribute-aware caching of directional ray quantities. The CHERT framework is evaluated through a series of numerical investigations. A brief nodal-scale benchmark identifies qP as a control mode, while qSV remains the stringent validation test for the selection of Hamiltonian solvers and optimizers. A field-scale test then benchmarks CHERT against the modified shortest-path method (MSPM), which is governed by the same g*-Hamiltonian. In the controlled benchmark, CHERT achieves substantially faster runtime than MSPM, with qP and qSV results that are effectively indistinguishable. Angular coverage and nodewise tests between CHERT and MSPM indicate that there is no missing acceptable local directions or path misselection. A modified Marmousi2 application demonstrates the application of CHERT in a 2D synthetic model, while a layered 3D computation demonstrates its potential for extension to viscoelastic anisotropic media. Additionally, the ablation analysis shows that a 1° sampled ray-direction step preserves qSV behavior in the evaluated configuration, while overly coarse cache-class grouping remains insufficient for the Marmousi2 case. These findings demonstrate that CHERT makes the g*-Hamiltonian kernel practical for model-scale first-arrival ray tracing while preserving the sampled directional ray-quantity manifold in viscoelastic anisotropic media.

## INTRODUCTION

The Earth's subsurface rarely behaves as a purely acoustic or elastic medium. Anisotropy and heterogeneity affect wave propagation, making the wavefield both directionally and spatially dependent. It also exhibits viscoelastic behavior that attenuates the energy of propagating waves, increasing the complexity of the propagation mechanism (Aki and Richards, 1980; Thomsen, 1986; Carcione, 2022). Under these conditions, seismic ray tracing remains an important tool in seismic imaging, tomography, ray-based amplitude analysis, and migration, as it translates the high-frequency asymptotics of wave propagation into physically permissible wavefronts, traveltime fields, and raypaths (Červený and Brown, 2001).

Classical ray tracing, based on the method of characteristics, provides a direct kinematic and dynamic framework for high-frequency wave propagation via ray trajectories and quantities. These methods include fundamental analytical and perturbation-based developments for characterizing anisotropic media and dynamic ray attributes (Červený and Brown, 2001; Courant and Hilbert, 2008; Huang and Greenhalgh, 2018). Nonetheless, their application, based on shooting and bending methods, generally relies on smoothly varying medium properties and can become unstable in the presence of sharp boundaries, discontinuities, and qSV cusps. This challenge is further compounded in viscoelastic anisotropic media (VEAM), where the energy velocity and associated ray quantities are complex-valued and depend on both a homogeneous slowness-angle solution and wave mode (Vavryčuk, 2007a, 2008a, 2012; Wang, 2014). Among the body wave modes, the qSV wave is particularly demanding because the presence of cusps, triplications, and attenuation can substantially alter the local wavefront configuration. Meanwhile, the reliable computation of complex energy velocity for all body waves (qP, qSV, and qSH) demands a search over local slowness-angle pairs that satisfy the homogeneity condition between the real and imaginary parts of the complex energy velocity (Li et al., 2020; Wu et al., 2021a; Wang et al., 2024; Zhou et al., 2024a, 2024b).

Grid-based shortest-path techniques remain attractive in this context, as they are well-suited for first-arrival computation in complex medium conditions such as anisotropy, sharp contrasts, and complex structures, without requiring shooting or bending evaluation on every candidate ray segment. In these techniques, the medium is discretized into a gridded network of nodes and cells to recover traveltimes and raypaths using a Dijkstra-type shortest-path search through the gridded network (Dijkstra, 1959). Notable developments have been achieved in 2D (Cao and Greenhalgh, 1993; Zhou and Greenhalgh, 2005, 2006) and 3D (Bai et al., 2007, 2016) anisotropic media, resulting in practical first-arrival ray tracing methods applicable to strongly heterogeneous anisotropic settings. Subsequently, several extensions to VEAM have incorporated ray-quantity solvers into various MSPM algorithms (Li et al., 2020; Wu et al., 2021b). However, field-scale modeling remains constrained by approximate solvers, smooth-gradient model parameterization, iterative local evaluations, and sensitivity to strong parameter contrasts or along layer boundaries. Furthermore, existing methods rely heavily on primary or corner nodes for model parameterization and for evaluating ray quantities, while treating the remaining nodes as auxiliary points. As a result, the field-scale propagation remains limited by interpolation or reconstruction of nodal ray quantities for non-corner nodes and repeated nodal Hamiltonian evaluations.

The main unresolved gaps remain at the model scale. The g*-Hamiltonian solver for ray quantities has improved the computation of local ray quantities, specifically for qSV wavefront and ray quantities affected by cusps or triplications (Zhou et al., 2024). However, directly integrating this solver with existing shortest-path-based methods remains expensive, as the solver must be called repeatedly at each node during node-to-node updates. This bottleneck becomes more pronounced for qSV waves, as cusp-related challenges, branch sensitivity, and subtle variations in attenuation can render ray-quantity computations highly sensitive to the ray-direction solution (Zhou et al., 2024b; Liu, 2025). Another limitation arises at the nodal-discretization level. In conventional shortest-path methods, ray directions within a cell are evaluated only at primary nodes and then interpolated to secondary nodes. However, discretized grid connectivity imposes additional constraints, such as limited angular resolution. Consequently, the sampled directional information of ray quantities can become constrained by node roles, interpolation accuracy, and sampling density. This limitation is not purely a fundamental physics issue; rather, it reflects the representation of how the nodal ray quantity manifold is translated into field-scale path search.

To address these limitations, we introduce CHERT for viscoelastic anisotropic media that retains the shortest-path framework while restructuring the integration of the ray-quantity solver. Instead of repeatedly solving the Hamiltonian problem at each candidate node, CHERT precomputes and stores directional ray quantities in indexed caches classified by the wave mode, medium's geometric and rock properties, including layer, facies, lithology, and symmetry-axis orientation. During node-to-node updates, each node determines its cache class and queries the ray quantities associated with the cached ray direction that closely matches the candidate path direction.

The main contributions are fourfold. First, CHERT introduces a cache-enhanced ray-tracing framework that decouples ray-quantity evaluation from field-scale modeling. Second, it defines a cache organization as a directional parameterization based on wave mode and cache class, enabling the benchmarking of g-Hamiltonian and g*-Hamiltonian kernel caches under the same geometric setting. Third, it compares the performance of CHERT with that of the MSPM in VEAM, which uses g*-Hamiltonian kernels. Fourth, it validates the framework through a controlled benchmark, angular coverage comparison at the node scale, modified Marmousi2 ray tracing, an extension to 3D TTI media, and two ablation studies.

We keep the kernel assessment concise, focusing only on the key aspects necessary to highlight the CHERT application and avoiding the scope of a full kernel development and benchmarking study (Zhou and Greenhalgh, 2004). Hence, the present work emphasizes the effectiveness of addressing qSV cusps and the importance of selecting an optimizer. Other ray-tracing techniques, namely, fast marching, eikonal, and angle-shooting-bending methods, remain important across a wide range of studies (Vidale, 1988, 1990; Rawlinson and Sambridge, 2004; Sethian, 2006), yet they do not address the same kernel benchmarking objectives considered in the present numerical investigation.

In this paper, we benchmark two complex energy velocity formulas, called the g-Hamiltonian and the g*-Hamiltonian, to review the best local kernel. The field-scale benchmark is then conducted between CHERT and MSPM in 2D VEAM under the same g*-Hamiltonian. We demonstrate the advantages of CHERT methods with g*-Hamiltonian kernel compared to the conventional MSPM, give computational examples applied to a practical rock sample, validate CHERT's performance under pointwise truth and angular coverage test, and provide CHERT's practical implementation on modified Marmousi2 and 3D viscoelastic TTI. We conduct two ablation studies to support the practical configuration of the CHERT's framework. Finally, we draw some general conclusions from these numerical experiments.

# THEORETICAL BACKGROUND

## Viscoelastic anisotropic local framework

In moderately heterogeneous viscoelastic anisotropic media without considering an external source, the equation of motion in the frequency domain (Hanyga and Seredyńska, 2000; Červený and Brown, 2001; Carcione, 2022) is written as,

$$\rho\omega^2\boldsymbol{u}_j + \rho a_{ijkl}\partial_{li}\boldsymbol{u}_k = 0 \tag{1}$$

where $\rho(\mathrm{x})$ denotes the density, $\omega$ is the angular frequency, $\boldsymbol{u}_j(\mathrm{x},\omega)$ is the $j^{th}$ displacement vector component, and $a_{ijkl}(\mathrm{x},\omega)$ denotes the density-normalized complex moduli and is expressed as,

$$a_{ijkl} = a_{ijkl}^{(R)} - i a_{ijkl}^{(I)} \tag{2}$$

Here, $a_{ijkl}^{(R)}$ and $a_{ijkl}^{(I)}$ represent the real and imaginary parts of the stiffness, respectively, where the real part represents the density-normalized stiffness modulus, while the imaginary part describes the dissipation and attenuation. We use the density-normalized modulus not only to keep the Hamiltonian expression concise but also to preserve a direct physical interpretation as the squared velocity. The time-harmonic ansatz of this medium (Červený and Brown, 2001; Vavryčuk, 2007b) is expressed through the following high-frequency asymptotic formula,

$$\boldsymbol{u}_i = \mathbf{A}_i e^{\mathrm{i}\omega(\tau - \mathbf{p}_i \cdot \mathrm{x}_i)} = |\mathbf{A}_i| \mathbf{g} e^{\mathrm{i}\omega(\tau - \mathbf{p}_i \cdot \mathrm{x}_i)} \tag{3}$$

where $\boldsymbol{u}_i$, $\mathbf{A}_i$, and $\boldsymbol{p}_i$ are the complex-valued displacement vector, ray amplitude, and slowness vector, respectively**,** while $\mathbf{g}$ is the complex-valued polarization vector (eigenvector) and $\boldsymbol{\tau}$ is the complex-valued traveltime. Substituting equation 3 into equation 1 yields the Christoffel equation,

$$\left[\boldsymbol{\Gamma}_{jk}(\mathrm{x}, \mathbf{p}) - \delta_{jk}\right] \mathbf{g}_k = 0, \;\; \boldsymbol{\Gamma}_{jk}(\mathrm{x}, \mathbf{p}) = a_{ijkl} \mathbf{p}_l \mathbf{p}_i \tag{4}$$

where $\mathbf{p}$ is the complex slowness vector, which can be written in terms of the complex-valued phase velocities $c$ and the complex slowness direction vector $\mathbf{n}$ as,

$$\mathbf{p} = \frac{\mathbf{n}}{c} \tag{5}$$

In VEAM, the complex slowness direction vector is given as (Hanyga and Seredyńska, 2000),

$$\mathbf{n} = \begin{bmatrix} \sin(\theta + i\vartheta)\cos(\varphi + i\psi) \\ \sin(\theta + i\vartheta)\sin(\varphi + i\psi) \\ \cos(\theta + i\vartheta) \end{bmatrix} = \mathbf{n}^{(R)} + i\mathbf{n}^{(I)} \tag{6}$$

where $(\theta, \vartheta)$ and $(\varphi, \psi)$ are inclination and azimuthal angle pairs of the slowness vector, respectively, in 3D space with respect to the local symmetry-axis frame (Zhou et al., 2024b). In the viscoelastic VTI medium, the azimuth angles are zero ($\vartheta = \psi = 0$), equation 6 is simplified as follows,

$$\mathbf{n} = [\sin(\theta + i\vartheta), 0, \cos(\theta + i\vartheta)] \tag{7}$$

**Hamiltonian Kernel: Ray Quantities and Complex Energy Velocity**

In VEAM, the definition of velocity is described as (Vavryčuk, 2007b, 2007a, 2008b),

$$\mathbf{v} = \frac{d\mathrm{x}}{d\tau} = \frac{d\mathrm{x}}{d\tau^{(R)} + d\tau^{(I)}} = \mathbf{v}^{(R)} + i\mathbf{v}^{(I)} \tag{8}$$

where $\mathbf{v}$ is the complex energy velocity, and $\boldsymbol{\tau}$ is the complex traveltime, both having real and imaginary parts. Their real and imaginary parts govern the propagation velocity and attenuation, respectively. The complex energy velocity $\mathbf{v}$ is the central foundation for the ray tracing framework and has the homogeneity condition that is written as follows (Zhou et al., 2024b),

$$\mathbf{v}^{(R)} = \frac{1}{1+\alpha^2}\frac{d\mathrm{x}}{d\tau^{(R)}}, \qquad \mathbf{v}^{(I)} = \frac{\alpha}{1+\alpha^2}\frac{d\mathrm{x}}{d\tau^{(R)}} \tag{9}$$

Equation 9 states that the real and imaginary parts of the complex energy velocity are collinear or homogeneous $\left(\mathbf{v}^{(R)} = \alpha\mathbf{v}^{(I)}\right)$, where $\alpha$ is a real-valued proportionality factor ($\alpha = -\frac{d\tau^{(I)}}{d\tau^{(R)}}$). This proportionality condition implies an acceptable homogeneous complex energy velocity solution. When the condition is satisfied, the real and imaginary parts share the same ray direction. Thus, $\mathbf{v}$ is called the homogeneous complex energy velocity vector (Vavryčuk, 2007a; Carcione, 2022), also known as the group velocity in elastic media. Let $\hat{\mathbf{r}}^{(R)}$ and $\hat{\mathbf{r}}^{(I)}$ denote the ray direction vectors of $\mathbf{v}^{(R)} = v^{(R)}\hat{\mathbf{r}}^{(R)}$ and $\mathbf{v}^{(I)} = v^{(I)}\hat{\mathbf{r}}^{(I)}$. Then, the traveltime perturbation can be expressed as,

$$d\tau = \frac{ds}{v.\hat{\mathbf{r}}} = \frac{ds}{[v^{(R)}(\hat{\mathbf{r}}^{(R)}.\hat{\mathbf{r}}) + iv^{(I)}(\hat{\mathbf{r}}^{(I)}.\hat{\mathbf{r}})]} = \frac{ds}{V^{Ray}} + iA^{Ray}ds, \tag{10}$$

where the ray velocity $V^{Ray}$ , attenuation $A^{Ray}$, and $Q^{Ray}$ are derived from homogeneous complex energy velocity as follows,

$$V^{Ray} = \frac{\left[v^{(R)}\right]^2 + \left[v^{(I)}\right]^2}{v^{(R)}}, \qquad A^{Ray} = \frac{-v^{(I)}}{[v^{(R)}]^2 + [v^{(I)}]^2}, Q^{Ray} = -\frac{\mathrm{Re}[v^2]}{\mathrm{Im}[v^2]} \tag{11}$$

The ray velocity $V^{Ray}$ corresponds to the propagation velocity along a ray, while $A^{ray}$ represents the ray attenuation that describes amplitude decay along a ray, and $Q^{ray}$ characterizes the ray quality factor. These parameters embody the ray quantities that characterize wave propagation along a ray (Vavryčuk, 2007a, 2007b, 2008b). According to equation 10, $V^{Ray}$ and $A^{Ray}$ affect traveltime and subsequent raypaths by shaping the traveltime perturbation, which serves as the foundation of field-scale ray tracing.

The first notable study in real-space ray tracing was conducted by Vavryčuk (2007a), who presented the first implementation in VEAM for qP and qSH waves and introduced the term *complex energy velocity*. Here, the g-Hamiltonian formula is referred to as the original energy velocity (OEV) and is used as an important reference, especially for nodal modeling of qP and qSH waves. Based on the characteristic methods (Courant and Hilbert, 2008), the OEV is formulated as (Červený and Brown, 2001; Vavryčuk, 2007a)

$$\mathbf{v} = \frac{d\mathrm{x}_i}{d\tau} = \frac{1}{2}\frac{\partial G}{\partial \mathbf{p}_i} = \frac{\partial H}{\partial \mathbf{p}_i} = a_{ijkl}\mathbf{p}_l\mathbf{g}_j\mathbf{g}_k \tag{12}$$

Several kernel formulations have been proposed to enhance the qSV ray quantity computation by incorporating conjugate eigenvectors into the Hamiltonian functional (Wu et al., 2021b; Zhou et al., 2024b). Here, the g*-Hamiltonian formulation is utilized as a Hamiltonian kernel, which was constructed from a conjugate eigenvector (Zhou et al., 2024b). The Hamiltonian form can be formulated as,

$$H(\mathrm{x}, \mathbf{p}) = \frac{1}{2}\left(\mathbf{p}.\bar{\bar{\mathbf{\Gamma}}}\mathbf{p} - 1\right) \tag{13}$$

Constructing the g*-Hamiltonian formula that reads

$$\mathbf{v} = \frac{\partial H}{\partial \mathbf{p}} = \bar{\bar{\mathbf{\Gamma}}}\mathbf{p} + \frac{1}{2}\mathbf{g}^{*}.\,[\mathbf{\Gamma}(\mathbf{p}) - \mathbf{I}]\frac{\partial \mathbf{g}}{\partial \mathbf{p}} \tag{14}$$

Here, we refer to the g*-Hamiltonian formula as the modified energy velocity (MEV). The MEV formula has been tested and can handle the presence of complex cusps or triplication on the qSV wavefront, providing an admissible wavefront and ray quantities of qP, qSV, and qSH wave modes (Zhou et al., 2024a, 2024b). Both OEV and MEV are treated here as local kernel options for integration with CHERT in field-scale modeling.

**Fermat's (Variational) Principle**

This principle serves as an important link between the local-scale energy velocity and ray quantities, and the traveltime and raypath modeling at the larger scale. In general, anisotropic media can be modeled using characteristic-based wave propagation and Fermat variational equations (Bóna and Slawinski, 2003; Courant and Hilbert, 2008). The complex traveltime $\tau$ is defined along the raypath $R$ as

$$\tau = \int_{\mathrm{R}} \mathbf{p}\, d\mathrm{x} = \int_{\mathrm{R}} \frac{ds}{\mathbf{v}} \tag{15}$$

in which d$s = |d\boldsymbol{x}|$ is a small part of the raypath. Substituting the first term of equation 10 into 15 gives

$$\tau = \int_{\mathrm{R}} \frac{v^{(R)}}{[v^{(R)}]^2 + [v^{(I)}]^2}\,\mathrm{d}s + i\int_{\mathrm{R}} \frac{-v^{(I)}}{[v^{(R)}]^2 + [v^{(I)}]^2}\,\mathrm{d}s = \int_{\mathrm{R}} \frac{1}{V^{Ray}}\,\mathrm{d}s + i\int_{\mathrm{R}} A^{Ray}\mathrm{d}s \tag{16}$$

Perturbing both sides of equation 16 results in

$$\delta\tau(\mathrm{x}_{AB}) = \delta\left(\int_{\mathrm{A}}^{\mathrm{B}} \frac{1}{V^{Ray}} ds + i\int_{\mathrm{A}}^{\mathrm{B}} A^{Ray} ds\right) = \delta\left(\int_{\mathrm{A}}^{\mathrm{B}} \frac{1}{V^{Ray}} ds\right) + i\delta\left(\int_{\mathrm{A}}^{\mathrm{B}} A^{Ray} ds\right) = 0 \quad (17)$$

Fermat's variational principle states that the physical raypath should be the one that yields the integral of traveltime at the stationary state $\delta\tau(\mathrm{x}_{AB}) = 0$ (Zhou and Greenhalgh, 2005, 2006). Based on this principle, the shortest raypaths from a source to a receiver must be the minima among all plausible paths. Consequently, the traveltime at points A to B is then constructed as

$$\tau_{AB}(\mathrm{x}) = \min\left[\int_{\mathrm{B}}^{\mathrm{A}} \frac{1}{V^{Ray}(\mathrm{x}, \hat{\mathbf{r}}_{AB})} \mathrm{d}s + i\int_{\mathrm{B}}^{\mathrm{A}} A^{Ray}(\mathrm{x}, \hat{\mathbf{r}}_{AB}) \mathrm{d}s\,, \forall\, \mathrm{x}_A \in \Omega_B\right] \quad (18)$$

while the subsequent raypath is formulated as

$$\mathcal{R}_{AB}^{(p)} = \arg \min_{\mathbf{x}_A \in \Omega_B} \left\{\int_{\mathrm{B}}^{\mathrm{A}} \frac{1}{V^{Ray}(\mathrm{x}, \hat{\mathbf{r}}_{AB})} ds\right\},\ \mathcal{R}_{AB}^{(a)} = \arg \min_{\mathbf{x}_A \in \Omega_B} \left\{\int_{\mathrm{B}}^{\mathrm{A}} A^{Ray}(\mathrm{x}, \hat{\mathbf{r}}_{AB}) ds\right\} \quad (19)$$

in which $\mathcal{R}_{AB}^{(p)}$ is the propagation raypath, implying paths along maximum propagation energy, and the traveltime is minimum (or stationary) while $\mathcal{R}_{AB}^{(a)}$ is the attenuation in raypath units describing the trajectories where the energy damping is at maximum.

In the continuous medium, wave propagation and attenuation are variational. Once the medium is discretized into nodes and cells, the admissible traveltime and raypath must be determined by the minimum traveltime between two possible nodes based on the related ray quantities. The main challenge is how to preserve these quantities across all nodes while still accounting for the physics, with rapid computational runtime. If a local Hamiltonian solver must be solved on every node-to-node update, then the raypath tracing must be expensive. However, if the local Hamiltonian solver must be solved only on certain nodes while the other nodes' quantities are interpolated, it can make the available directional ray quantities depend on node role and nodal density.

## CHERT METHODOLOGY

This section explains how CHERT uses the ray quantities generated by the Hamiltonian solvers (MEV or OEV) in its computations. CHERT is formulated to address problems with existing shortest-path-based methods, such as the repeated nodal Hamiltonian evaluations and directional ray-quantity interpolation of directional ray quantities at non-corner nodes. It then reorganizes how the nodal directional ray quantities are made available to model traveltimes and drive raypaths. The key innovation here is the introduction of a cache architecture that separates the node-scale solver from field-scale path determination. The cache allows us to reformulate how the sampled directional information from a local kernel is preserved, retrieved, and reutilized in the field-scale evaluation during node-to-node updates.

### Method Overview and Design Principles

In general, CHERT is a cache-augmented shortest-path framework designed for viscoelastic anisotropic media. The medium is discretized into a gridded nodal network as the discrete ray-tracing framework, in which geological structures, layer boundaries, receivers, and sources are comprised in the network and assigned within the nodes. The main objective is to preserve the sampled directional manifold of Hamiltonian-based ray-quantity information while avoiding iterative precomputation at every candidate node. Two distinct stages separate the full framework. First, the ray quantities are parameterized and stored in the cache based on predetermined facies, layers, or TTI symmetry-axis orientation. Second, propagation and attenuation in traveltimes and raypaths are calculated by retrieving the relevant cached manifolds at each node.

This separation is the central concept of CHERT. In conventional shortest-path applications, such as MSPM, a localized table is regenerated once a cell is active, using ray quantities associated with its corner nodes. All of these corner nodes require a fresh Hamiltonian evaluation to confine the ray quantities. Ray quantities at non-corner nodes are then provided through corner-derived reconstruction. Within CHERT, the evaluation is precomputed once per wave mode, facies, and sampled direction before the path updates, allowing nodes along the candidate paths to query the resulting directional quantities directly across the entire grid network. The cache-augmented formulation remains consistent with the shortest-path logic, in which path selection and cell relaxation are driven by discrete minimization, while nodal parameters are retrieved from cache.

CHERT's discrete minimization is based on Fermat's and Huygens' principle (Musgrave, 1970). The former yields the variational fundamental of the shortest raypath described in the previous section, while the latter characterizes the propagation of the wave by supposing that each point on a wavefront serves as an elementary source of secondary wavelets. The resulting wavefront is determined by the accumulation of effects from all the points on the previous wavefront. Under the proposed formulation, the raypath update follows Fermat's principle by identifying the minimum traveltime among admissible node-to-node raypath segments. The cell-wise marching order follows Huygens-type wavefront expansion for each wave mode, in which adjacent cells are visited according to the minimum tentative traveltime. CHERT will be presented hereafter, following these foundations.

**Grid and Nodal Ray-Network Representation**

The 2D or 3D viscoelastic anisotropic media are discretized into a gridded model constructed with a number of non-overlapping cells (2D) or blocks (3D) $\mathcal{N}_b$ . Each discrete element $\mathcal{N}_b$ comprises a finite set of non-overlapping nodes $\mathcal{N}_n$ along the cell edges. Each cell and node is associated with a specified cache class according to $l^{th}$ layer, facies, structural class, or rock type. Each node serves as a possible point on a raypath or segment within a cell. All possible raypaths from a source are reconstructed through this discretized network to all receivers.

Based on the block or cell and node IDs, at node $n$ from source $s$, the complex traveltime of wave mode $m$ is individually denoted by

$$\tau_n^{(s,m)} = \tau_n^{(s,m,R)} + i\tau_n^{(s,m,I)} \tag{22}$$

Each node eventually retains only the propagation and attenuation in traveltime, along with the preceding node indices needed to model propagation and attenuation in raypaths. For a block-level marching, the traveltime at block $b$ from source $s$ denotes

$$T_b^{(s,m)} = T_b^{(s,m,R)} + iT_b^{(s,m,I)} \tag{23}$$

where each part corresponds to the smallest node traveltime in a block or cell as

$$T_b^{(s,m,R)} = \min_{n\in\mathcal{N}(b)} \tau_n^{(s,m,R)}, \quad T_b^{(s,m,I)} = \min_{n\in\mathcal{N}(b)} \tau_n^{(s,m,I)} \tag{24}$$

The attenuation and propagation parts will individually drive the numerical framework to determine the following block and cell within the adjacent blocks using a Dijkstra-like expansion algorithm, selecting the blocks with the smallest current traveltime. In addition, the corresponding propagation and attenuation ray trajectories from the source to the receiver of wave mode $m$ are expressed as a sequence of node indices

$$\mathcal{R}_{s,r}^{(s,m,p)} = \left(i_1^{(s,m,p)}, i_2^{(s,m,p)}, \dots, i_r^{(s,m,p)}\right), \qquad \mathcal{R}_{s,r}^{(s,m,a)} = \left(i_1^{(s,m,a)}, i_2^{(s,m,a)}, \dots, i_r^{(s,m,a)}\right) \quad (25)$$

where $i$ is the node index passed through the propagation and attenuation raypath from the first index $i_1$ to receiver $i_r$ of wave mode $m$. For this reason, each node preserves the minimum real and imaginary traveltime and logs the predecessor node that yields the minimum traveltime**.** This gridded ray-network transfers the local physics information into field-scale ray tracing through discrete relaxation. Through cache lookup, the suitable directional ray quantities are extracted for each node-to-node path candidate and utilized to allocate the propagation or attenuation increment. At this point, it is important to distinguish the discretized-medium representation from the directional information associated with it. The gridded network determines where propagation and attenuation may occur, while the cache defines how stored ray quantities are supplied to model the raypaths and traveltime fields. This numerical setup enables CHERT to be modular and extensible.

In the proposed discretized network, nodes are not classified as primary (corner) or secondary nodes. The proposed method does not assign a specific physical role solely to certain nodes. All nodes can query the cache for relevant quantities and parameters needed while selecting the possible trajectories. In addition, by enhancing the medium to include attenuation and direction dependencies, the number of parameters per node or block increases considerably. Consequently, the grid may represent numerous multiparameter values, i.e., the numbers of stiffness moduli and quality factors, rock densities for each lithology or layer, and ray quantities. These parameters are directionally and/or spatially based on the ray direction, attenuation level (amplitude decay intensity), lithologies, or facies. Such a strategy would clearly increase the computational cost, especially if it is performed on every node, e.g., corner nodes. Likewise, the effect of perturbing the number of nodes is examined.

**Directional Cache Construction**

Let $k$ denote the cache class, which is the characterized group for ray quantities and medium parameters tabulation. It corresponds to a facies group, a geological layer, a structural unit, a rock type, a TTI symmetry-axis orientation, or even a cellwise grouping (Table 1). The same tabulation applies in 3D, with additional inclination and azimuthal TTI orientation that define the cache class. The number of cache classes is fundamental, as it determines how reliable the stored quantities are for a given part of the medium. All cells and nodes assigned to the same $c$ will share the same precomputed ray quantities and material parameters for each wave mode and sampled ray direction. Based on these classes, we define the cache $\mathcal{C}$ based on the cache class $k$, wave mode $m$, and sampled slowness-direction angles $\theta$ to store specific modulus, ray quantities, and directions as follows,

$$\mathcal{C}(m,k,\theta) \coloneqq \left(c_{ijkl}(k),\ Q_{ijkl}(k), \rho(k),\ V^{Ray}(m,\theta),\ A^{Ray}(m,\theta), \hat{\mathbf{r}}(m,\theta)\right) \tag{26}$$

The cache $\mathcal{C}$ serves as a lookup table for storing multiple variables (Figure 1, Table 1). If the medium is transversely isotropic, $\theta$ ranges from $0^0$ to $90^0$ that could be sampled at $1^0$ or $0.5^0$. A central cache-configuration question is whether there is a trade-off between fine-tuning the number of classes and computational runtime. The cache index $\theta$ denotes the sampled ray direction angle. For each $\theta$ value, the Hamiltonian kernel solves the homogeneity require on complex energy velocity by selecting the acceptable imaginary slowness angle and storing the corresponding ray quantities. In VTI, $\theta$ is measured in the global symmetry-axis frame, ranging from $0^0$ to $90^0$, while in TTI, the same range is measured after rotating the local coordinate frame into the global frame associated with the local orientation angles. While the cache organizes the precomputed directional ray quantities, the number of nodes determines the grid network performance to provide reliable nodal angular coverage that matches the precomputed ray directions $\hat{\mathbf{r}}$ on raypath selection.

TABLE 1. An example of a cache architecture in CHERT in viscoelastic VTI. The modulus, ray quantities, and directions are tied to specific classes (layers) and wave modes.

| | Cache Class: Layer | | | | | | | | |
|---|---|---|---|---|---|---|---|---|---|
| | $k_1$ | | | $k_2$ | | | $k_3$ | | |
| Wave mode | qP | qSV | qSH | qP | qSV | qSH | qP | qSV | qSH |
| Slowness-direction angles | $TI\ (0{:}90)$ | | | | | | | | |
| Stored Variables | $c_{ijkl}(1), Q_{ijkl}(1), \rho(1)$ | | | $c_{ijkl}(2), Q_{ijkl}(2), \rho(2)$ | | | $c_{ijkl}(3), Q_{ijkl}(3), \rho(3)$ | | |
| | $V_{qP}^{Ray}(TI)$ | $V_{qSV}^{Ray}(TI)$ | $V_{qSH}^{Ray}(TI)$ | $V_{qP}^{Ray}(TI)$ | $V_{qSV}^{Ray}(TI)$ | $V_{qSH}^{Ray}(TI)$ | $V_{qP}^{Ray}(TI)$ | $V_{qSV}^{Ray}(TI)$ | $V_{qSH}^{Ray}(TI)$ |
| | $A_{qP}^{Ray}(TI)$ | $A_{qSV}^{Ray}(TI)$ | $A_{qSH}^{Ray}(TI)$ | $A_{qP}^{Ray}(TI)$ | $A_{qSV}^{Ray}(TI)$ | $A_{qSH}^{Ray}(TI)$ | $A_{qP}^{Ray}(TI)$ | $A_{qSV}^{Ray}(TI)$ | $A_{qSH}^{Ray}(TI)$ |
| | $Q_{qP}^{Ray}(TI)$ | $Q_{qSV}^{Ray}(TI)$ | $Q_{qSH}^{Ray}(TI)$ | $Q_{qP}^{Ray}(TI)$ | $Q_{qSV}^{Ray}(TI)$ | $Q_{qSH}^{Ray}(TI)$ | $Q_{qP}^{Ray}(TI)$ | $Q_{qSV}^{Ray}(TI)$ | $Q_{qSH}^{Ray}(TI)$ |
| | $\hat{\mathbf{r}}_{qP}(TI)$ | $\hat{\mathbf{r}}_{qSV}(TI)$ | $\hat{\mathbf{r}}_{qSH}(TI)$ | $\hat{\mathbf{r}}_{qP}(TI)$ | $\hat{\mathbf{r}}_{qSV}(TI)$ | $\hat{\mathbf{r}}_{qSH}(TI)$ | $\hat{\mathbf{r}}_{qP}(TI)$ | $\hat{\mathbf{r}}_{qSV}(TI)$ | $\hat{\mathbf{r}}_{qSH}(TI)$ |

The cache maps any propagation direction between two nodes to the closest tabulated ray direction and reuses the associated ray quantities. This strategy separates the expensive ray-quantity computation (based on OEV or MEV) from the subsequent ray tracing, which is a table lookup operation. Hence, the method only needs to compute the multiparameter properties into ray quantities once per layer/facies and subsequently cache them, rather than iteratively computing the ray quantities on every node or corner node. The cache can be constructed in parallel or even before creating the discretized model, as long as the number of facies or layers is known. If the medium is isotropic or visco-isotropic, one may reduce the number of independent moduli and simplify the velocity formulation in terms of the complex group velocities of P- and S-waves, whereas for a non-attenuating medium, one may omit the quality factor element in the proposed ray-tracing workflow.

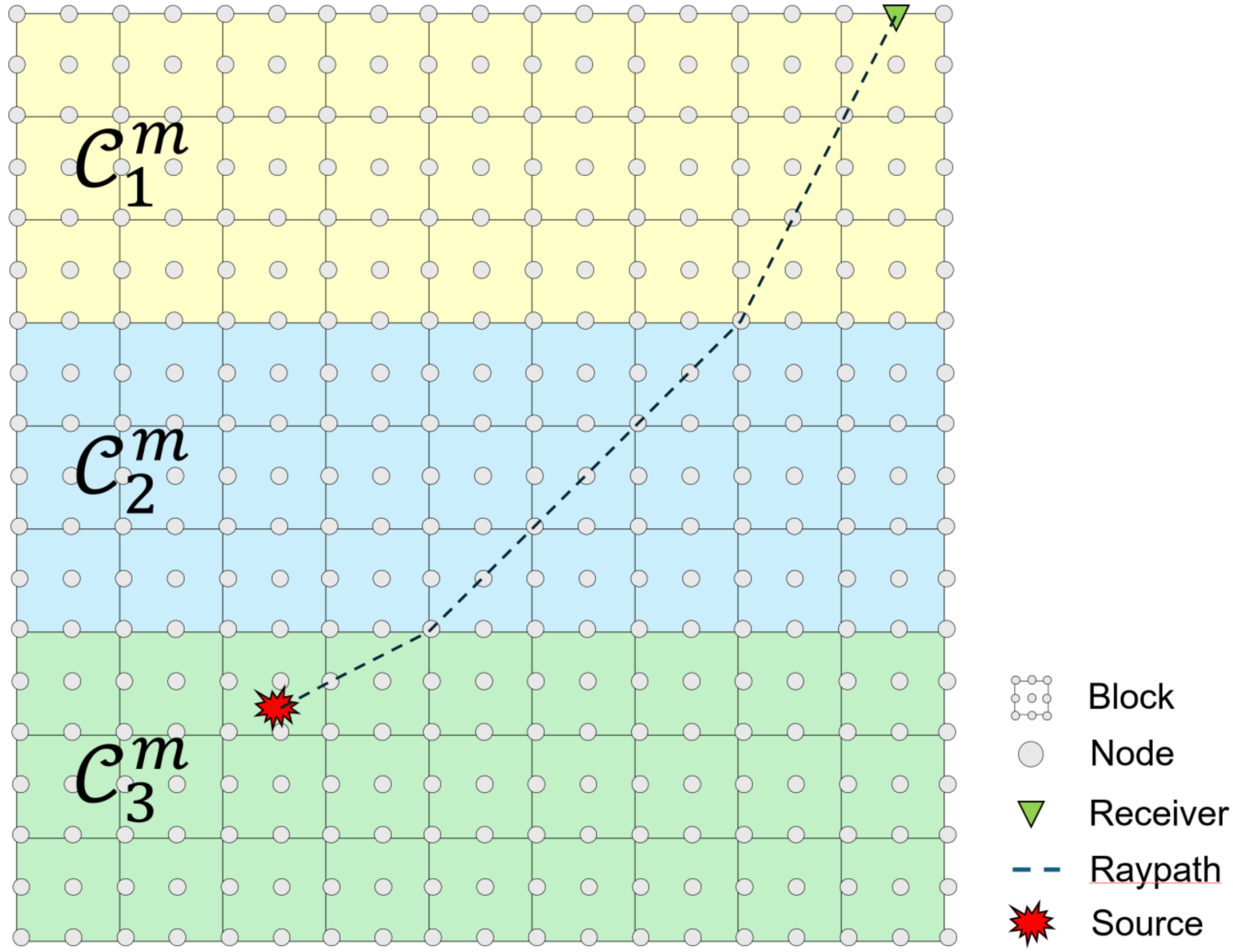

Figure 1. An example of a discretized and parameterized model in 2D with three classes corresponding 3 different layers containing the multiparameter variables $(\mathbf{a}_{\mathrm{ijkl}}, \mathrm{V}^{\mathrm{Ray}}, \mathrm{A}^{\mathrm{Ray}}, \hat{\mathrm{r}})$ of one m wave mode. Note that these caches will be recalled on every candidate node and are not placed into the grid, while all nodes (gray dots) serve equally as the possible path of the raypath (dashed black line) from a source (red star) and a receiver (green triangle).

**Traveltime and Raypath Algorithm**

After cache construction, for each source $s$ and wave mode $m$, the tentative minimum traveltime computation and the corresponding raypath reconstruction from the source to all locations (nodes) in the medium are preserved using a node-to-node local raypath update. To initiate ray tracing of a common shot gather, the traveltime in all nodes and blocks is set to infinity ($+\infty$). Meanwhile, on current-source nodes and blocks, the traveltime is zero. All other nodes and blocks are then considered as initially unreachable. All blocks with such traveltime conditions are initially labeled as non-visited, except for the source node and the block labeled visited.

The computation begins by identifying the minimum traveltime within the source block or cell, which will serve as the focal point of the raypath. The traveltime computation leverages (1) a block-wise marching scheme steered by a priority queue over blocks to update the overall block traveltime, and (2) a relaxation operator that imposes Fermat's principle inside each block, updating the internal nodes in that block. The subsequent raypath is modeled by backtracking along the recorded predecessor nodes that contain the neighboring node ID with the smallest traveltime update.

We use a priority queue $\mathscr{H}$, or heap, to determine the order in which to update the blocks, whose elements are block indices and traveltimes $\left(\left[b, T_k^{(s,m,R)}\right], \left[b, T_k^{(s,m,I)}\right]\right)$. The heap provides a queue for each block candidate or neighbor, ordered by priority, ensuring that the computation always executes the block with the smallest tentative traveltime. Once the heap stores the current block, it guarantees that, at every iteration, the next block computed is the one with the smallest tentative traveltime. We initialize the heap at the block source following the initialization setting $\mathcal{H} = \{(k_s, 0)\}$ and mark all cells as visited. Then, while the $\mathcal{H}$ is nonempty, any neighboring block from the source block with the minimum traveltime is explored and marked as the visited block(s). Inside the current source block, for each arbitrary pair of nodes, i.e., $i$ to $j$, the distance $\mathbf{d}_{ij}$ between them is considered as Euclidean distance, and a raypath candidate that reads

$$\mathbf{d}_{ij} = \hat{\mathbf{d}}_{ij}.\mathrm{d}_{ij} = \hat{\mathbf{d}}_{ij}\left|\left|\mathrm{x}_j - \mathrm{x}_i\right|\right|, \qquad \hat{\mathbf{d}}_{ij} = (\mathrm{x}_j - \mathrm{x}_i)/\left|\left|\mathrm{x}_j - \mathrm{x}_i\right|\right| \tag{27}$$

in which its unit vector $\hat{\mathbf{d}}_{ij}$ represents one candidate path direction between nodes $i$ and $j$ that matches the $\hat{\mathbf{r}}_{ij}$. The collection of $\hat{\mathbf{d}}$ from node $i$ to any arbitrary $j$ inside a cell corresponds to nodal angular coverage available for shortest-path identification. As the medium is discretized, this nodal angular coverage is controlled by the number of nodes within a cell or block to match the available sampled ray direction angles. Increasing nodal density enriches this angular coverage, whereas reducing nodal density reduces computational cost but may underresolve certain candidate directions and affect the path selection. This trade-off is particularly sensitive to qSV wave attenuation, especially in magnitude, as qSV wave magnitude varies only slightly or is highly nonlinear with the anisotropic angles. CHERT then has two practical parameters that must be carefully investigated and selected to achieve physical fidelity.

Let $l_i$ and $l_j$ be the layer indices of nodes $i$ and $j$, respectively. For the current $m$ wave mode computation, one may query the cache $\mathcal{C}_{l_i}^m$ and $\mathcal{C}_{l_j}^m$ to retrieve the associated $\hat{\mathbf{r}}_i(m,\theta)$, $\hat{\mathbf{r}}_j(m,\theta)$, $V_i^{Ray}(m,\theta)$, $V_j^{Ray}(m,\theta)$, $A_i^{Ray}(m,\theta)$ and $A_j^{Ray}(m,\theta)$. At each endpoint, the ray directions that best match $\hat{\mathbf{d}}_{ij}$ is selected for using a normalized dot product. The corresponding sampled ray direction angles $\theta_i$ and $\theta_j$ are then retrieved from precomputed tabulated data, which correspond to $V_i^{Ray}(m,\theta_i)$, $V_j^{Ray}(m,\theta_j)$, $A_i^{Ray}(m,\theta_i)$ and $A_j^{Ray}(m,\theta_j)$. Furthermore, the framework queries the segment-averaged ray quantities as

$$\bar{V}^{Ray}\left(m,\hat{\mathbf{d}}_{ij}\right)=\frac{1}{2}\left(V_i^{Ray}(m,\theta_i)+\ V_j^{Ray}\left(m,\theta_j\right)\right)$$
$$\bar{A}^{Ray}\left(m,\hat{\mathbf{d}}_{ij}\right)=\frac{1}{2}\left(A_i^{Ray}(m,\theta_i)+\ A_j^{Ray}\left(m,\theta_j\right)\right) \tag{28}$$

where the averaged ray velocity $\bar{V}^{Ray}\left(m,\hat{\mathbf{d}}_{ij}\right)$ and ray attenuation $A^{Ray}\left(m,\hat{\mathbf{d}}_{ij}\right)$ serve to build the traveltime increment along $\mathbf{d}_{ij}$ in which, for the propagation parameter, takes the form

$$\Delta\tau_{i\leftrightarrow j}^{(s,m,p)}=\frac{\left|\left|\mathrm{x}_j-\mathrm{x}_i\right|\right|}{\overline{\boldsymbol{V}}_{ij}^{Ray}\left(m,\hat{\mathbf{d}}_{ij}\right)} \tag{29}$$

while the attenuation parameter can be recast as

$$\Delta\tau_{i\leftrightarrow j}^{(s,m,a)}=\left|\left|\mathrm{x}_j-\mathrm{x}_i\right|\right|*\bar{A}^{Ray}\left(m,\hat{\mathbf{d}}_{ij}\right) \tag{30}$$

The complex traveltime at each node is then updated or “relaxed” in both directions, formulated as a relaxation step that follows

$$\tau_j^{(s,m,\odot)}\Leftarrow\min\left(\tau_j^{(s,m,\odot)},\tau_i^{(s,m,\odot)}+\Delta\tau_{i\leftrightarrow j}^{(s,m,\odot)}\right)$$
$$\tau_i^{(s,m,\odot)}\Leftarrow\min\left(\tau_i^{(s,m,\odot)},\tau_j^{(s,m,\odot)}+\Delta\tau_{i\leftrightarrow j}^{(s,m,\odot)}\right) \tag{31}$$

where, for conciseness, the $\odot$ denotes the real and imaginary part. Whenever an update is made, the predecessor node index is recorded and reset in the succeeding node, thereby representing a segment of the raypath. Propagation and attenuation are updated in parallel but remain computationally distinct, so that the computed traveltimes and back-traced trajectories can be analyzed separately. After determining the minimum traveltime by local relaxation at all nodes in a current source block, the smallest traveltime is assigned to the block traveltime that is given by

$$aT_{k_s}^{(s,m,\odot)}=\min_{j\in\mathcal{N}(k_s)}\tau_j^{(s,m,\odot)} \tag{32}$$

The computation continues to neighboring blocks until all unvisited blocks are visited and all traveltimes at the nodes are updated. Once the traveltime fields and preceding node indices are computed for all nodes and blocks, the raypaths, as a series of sequential node IDs, are reconstructed from the source node to all receiver nodes by backtracking from the recorded node indices at the receiver nodes to the source node. This backtracking process is done by the parent pointers $\left(\mathrm{par}^{(s,m,p)}, \mathrm{par}^{(s,m,a)}\right)$ at each node to refer to the previous nodes. Referring to the notation system of equations 31 and 32, the parent pointers are equivalently written as,

$$i_{D^{(m,\odot)}}^{(s,m,\odot)} = i_r, \qquad i_{l-1}^{(s,m,\odot)} = \mathrm{par}^{(s,m,\odot)}\left(i_l^{(s,m,\odot)}\right) \tag{33}$$

for $l = i_r, i_r - 1, \dots, i_s$ . The associated real-space coordinates along the propagation and attenuation paths may be expressed as

$$\mathrm{x}_q^{(s,m,\odot)} = \left(x_{i_q}^{(s,m,\odot)}, z_{i_q}^{(s,m,\odot)}\right), \qquad q = i_s, i_s + 1, \dots, i_r \tag{34}$$

These sequential points yield a piecewise computation of the continuous viscoelastic anisotropic raypaths for the first-arrival events in a 2-D model. After all paths from receiver nodes reach a source node, the computation is iteratively restarted to model the raypaths and traveltime fields for each source node location until all source nodes are considered.

Such computation is performed on individual wave modes and on separate components of propagation and attenuation. This segregation avoids mixing propagation and attenuation raypaths and traveltime values between each node and the complex part, which is important for achieving a clear characterization, especially when using the results as the initial model for multiparameter tomography and/or seismic migration. In addition, by tracing backward from each receiver node, the framework ensures that all raypaths reach all receivers.

**Computational Complexity**

The overall runtime costs of CHERT stem from the cache construction and field-scale raypath determination stages. The cache construction runtime is mainly affected by the number of classes, wave modes, and sampled local directions. Determining the raypath scales proportionally with the size of the overall medium, the number of blocks or cells, and the number of nodes, sources, and receivers. Nevertheless, compared with the conventional shortest-path-based method, CHERT replaces the expensive, repetitive nodal solvers with a computationally inexpensive, reusable cache. The simplicity of costs is the main advantage of CHERT for application in large, complex models and for its extension to 3D and TTI.

The framework design is primarily controlled by node density and cache-class characterization. Highly dense node space along edges preserves reliable, accurate results while increasing cache build elapsed time. A high number of classes enhances spatial representativeness but also yields slower computational time. These trade-offs are not merely practical details; rather, they directly affect the accuracy and uncertainty, especially for the qSV wave. Numerical investigations are conducted in the later sections.

**Tilted TI Geometry and Direction**

The framework is developed for both VTI and TTI media. Accordingly, the local direction must be depicted relative to the symmetry axis rather than only in the global frame. A VTI medium is then considered as a special case for a more general TTI configuration. For a TTI medium, the slowness direction follows the same constitutive law as a VTI medium but in a local coordinate frame whose symmetry axes are tilted relative to the global frame (Zhang and Zhou, 2018). This geometric extension allows local frame evaluations, which are then mapped back into global coordinates (Zhou and Greenhalgh, 2006). The rotating unit vectors are defined as follows,

$$\begin{aligned}\mathbf{e}_{x'} &= (\cos\theta_0\cos\phi_0)\mathbf{e}_x + (\cos\theta_0\sin\phi_0)\mathbf{e}_y - \sin\theta_0\mathbf{e}_z \\ \mathbf{e}_{y'} &= -\sin\phi_0\mathbf{e}_x + \cos\phi_0\mathbf{e}_y \\ \mathbf{e}_{z'} &= (\sin\theta_0\cos\phi_0)\mathbf{e}_x + (\sin\theta_0\sin\phi_0)\mathbf{e}_y + \cos\theta_0\mathbf{e}_z\end{aligned} \tag{35}$$

where $\mathbf{e}_{x'}, \mathbf{e}_{y'},$ and $\mathbf{e}_{z'}$ denote the rotated local basis vectors from the global Cartesian frame vector $(\mathbf{e}_x, \mathbf{e}_y, \mathbf{e}_z)$. The inclination angle $\theta_0$ and azimuthal angle $\phi_0$ correspond to the TTI angles associated with the vertical and horizontal rotations, respectively, that define the direction of the symmetry axis (Figure 2). For a 2D TTI case, $\phi_0 = 0$, equation 20 is reduced to

$$\begin{aligned}\mathbf{e}_{x'} &= (\cos\theta_0)\mathbf{e}_x - (\sin\theta_0)\mathbf{e}_z \\ \mathbf{e}_{z'} &= (\sin\theta_0)\mathbf{e}_x + (\cos\theta_0)\mathbf{e}_z\end{aligned} \tag{36}$$

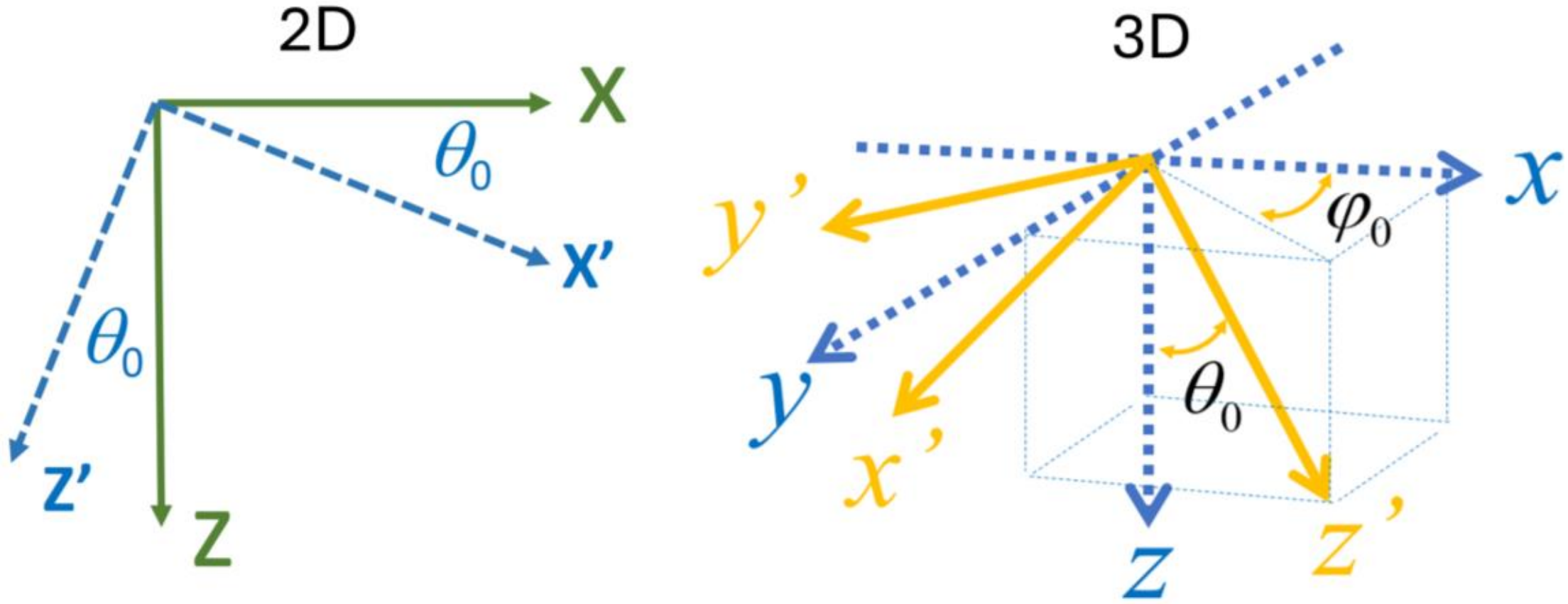

Figure 2. Symmetry axis rotation in 2D and 3D TTI frames, where the local frame (yellow and green solid lines) represents a VTI frame tilted at angles $(\theta_0, \varphi_0)$ relative to the global frame (blue-dashed lines).

# NUMERICAL EXPERIMENTS

The numerical investigation is organized in three tiers. A nodal kernel benchmark is first conducted to finalize the kernel and optimizer. Next, a controlled field-scale layered 2D viscoelastic VTI model is used to benchmark CHERT against the conventional shortest-path method (MSPM) baseline, using the same nodal kernel. Finally, CHERT is applied on a modified Marmousi2 model and a compact 3D TTI extension. To support the experiments, two ablation studies are conducted to identify reliable nodal density and grouping settings for cache building. Across all experiments, the reported parameters include the complex traveltime field, raypaths, runtime, and discrepancy metrics.

## Local kernel and optimizer benchmark

The first numerical experiment focuses on evaluating the local-physics options that need to be embedded in CHERT. The MEV and OEV serve as the physics-based kernel, with several tests on optimization functions, namely fminbnd, fminunc, and fsolve. These functions directly solve the slowness-angle pair needed to achieve the homogeneity condition using constrained searching (fminbnd) or unconstrained methods (fsolve and fminunc). For a comparative experiment, the modulus used the density modulus and quality factor of the siliciclastic rock tested in the g-Hamiltonian study, as shown in the first layer of Table 1 (Vavryčuk, 2007a). The study uses a wide range of searching angles $[-30^0 \leq \theta^I \leq 30^0]$ as a constraint for fminbnd.

The qP benchmark shows that the OEV and MEV remain coherent and produce stable results along the angular range on $V_{qP}^{ray}$, $A_{qP}^{ray}$, and $Q_{qP}^{ray}$, even though they have different smooth slowness angle-pairs (Figure 3). The optimization choice has only a negligible effect on controlling the final results. This effect is expected, as the qP and qSH waves should serve as controlling wave modes, with the main challenge in solving the physics on the nodal scale being the qSV wave.

In contrast, the qSV benchmark demonstrates several contrasting conditions. The OEV results for angle pairs and ray quantities are highly irregular, heavily dependent on the optimizer selection, and sensitive to changes in the VTI angle along the real slowness direction. The changes are highly abrupt in the range of $0^0 - 60^0$ angles on the ray quantities, while the $\theta^I_{qSV}$ shows noticeable oscillation (Figure 3). This condition is a prominent issue for the OEV, prompting several enhancements, including MEV. The MEV provides stable, smooth results, regardless of the optimization function used. The runtime results for optimization functions show that fsolve is the fastest (12.2s), followed by fminbnd (39.2s) and fminunc (57.1s). From all results, under this siliciclastic parameter, the range of $\theta^I$ is very small $[-2^0 \leq \theta^I \leq 0.5^0]$, indicating the challenging situation if the MEV or OEV must be executed using conventional iteration.

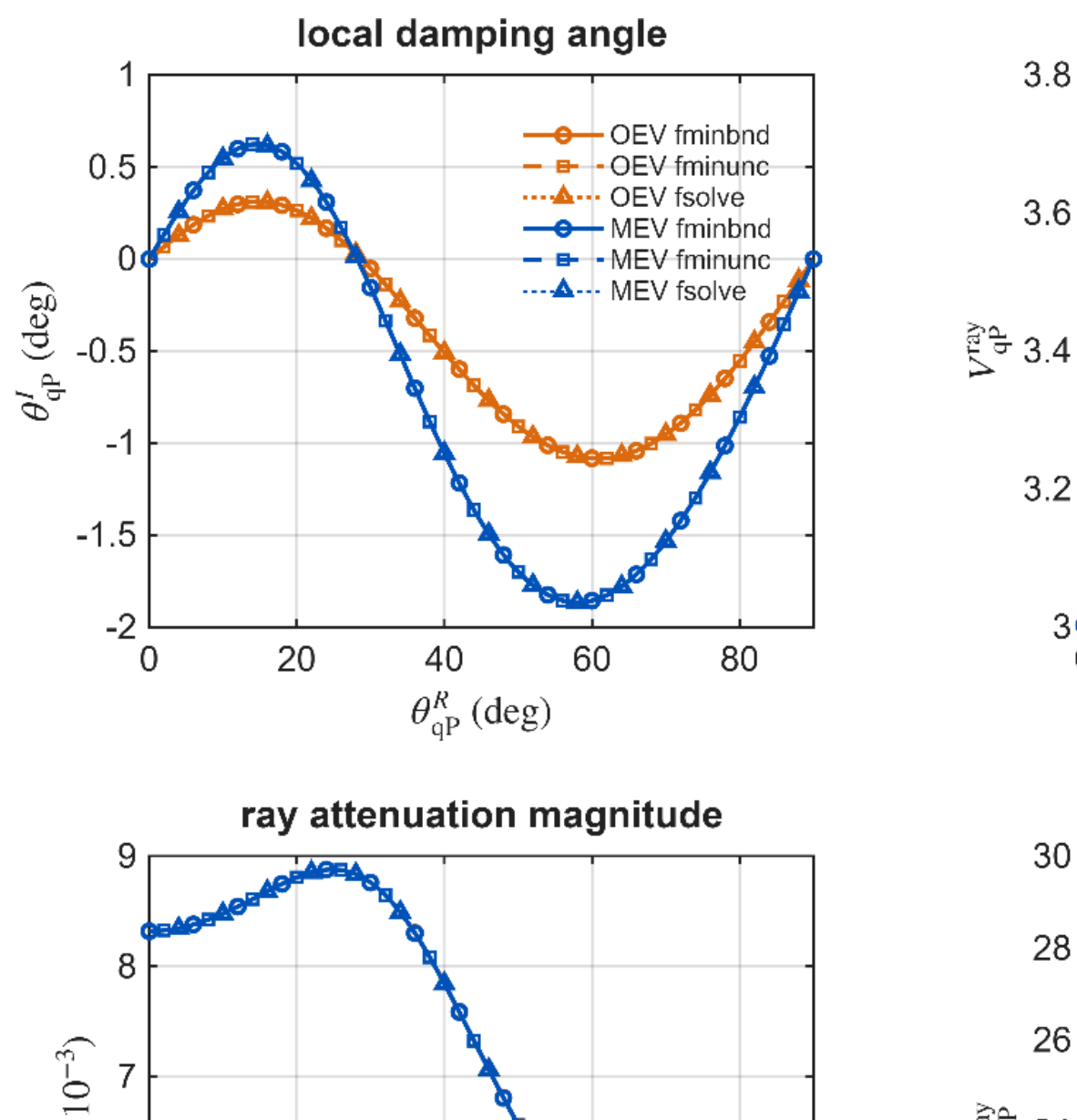
local damping angle
$\theta^I_{qP}$ (deg)
$\theta^R_{qP}$ (deg)
OEV fminbnd
OEV fminunc
OEV fsolve
MEV fminbnd
MEV fminunc
MEV fsolve

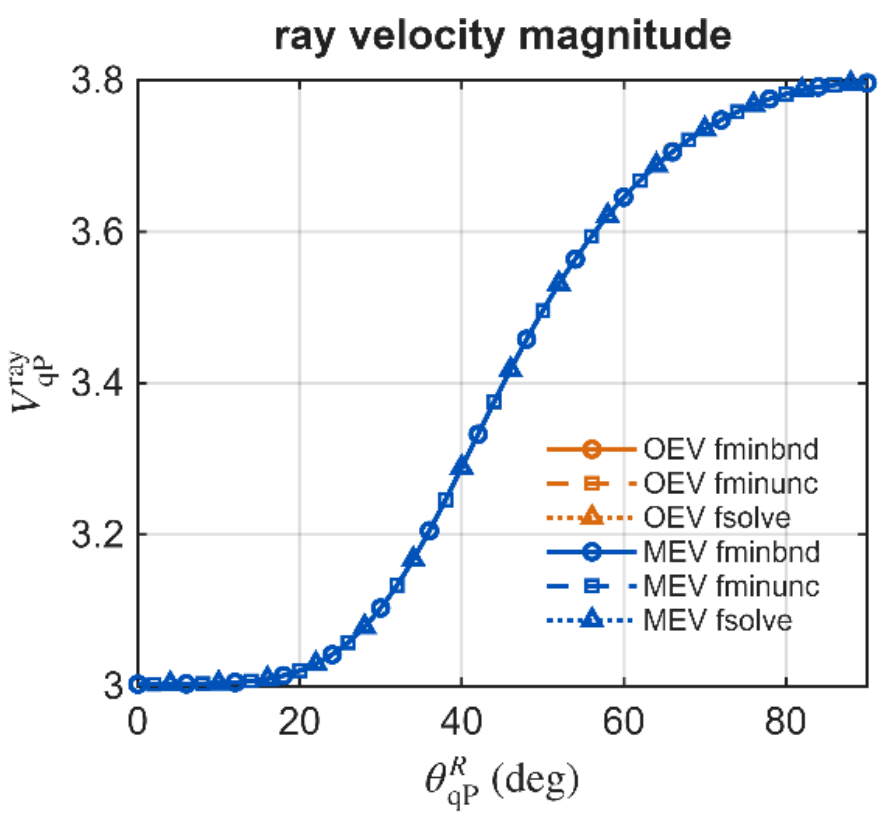
ray velocity magnitude
$V^{ray}_{qP}$
$\theta^R_{qP}$ (deg)
OEV fminbnd
OEV fminunc
OEV fsolve
MEV fminbnd
MEV fminunc
MEV fsolve

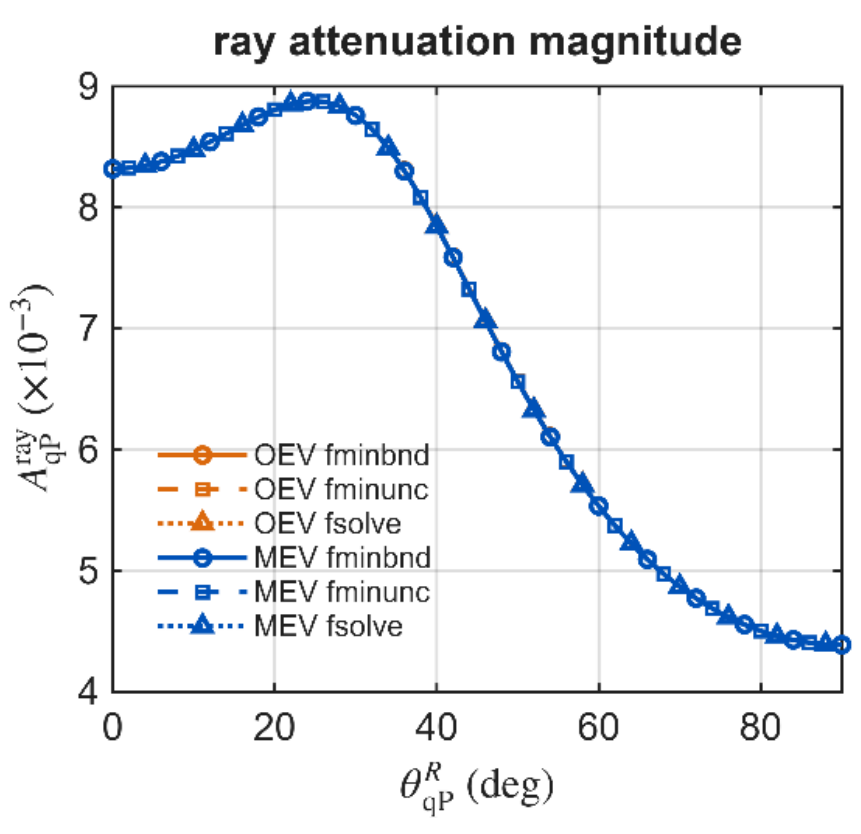
ray attenuation magnitude
$A^{ray}_{qP}$ ($\times 10^{-3}$)
$\theta^R_{qP}$ (deg)
OEV fminbnd
OEV fminunc
OEV fsolve
MEV fminbnd
MEV fminunc
MEV fsolve

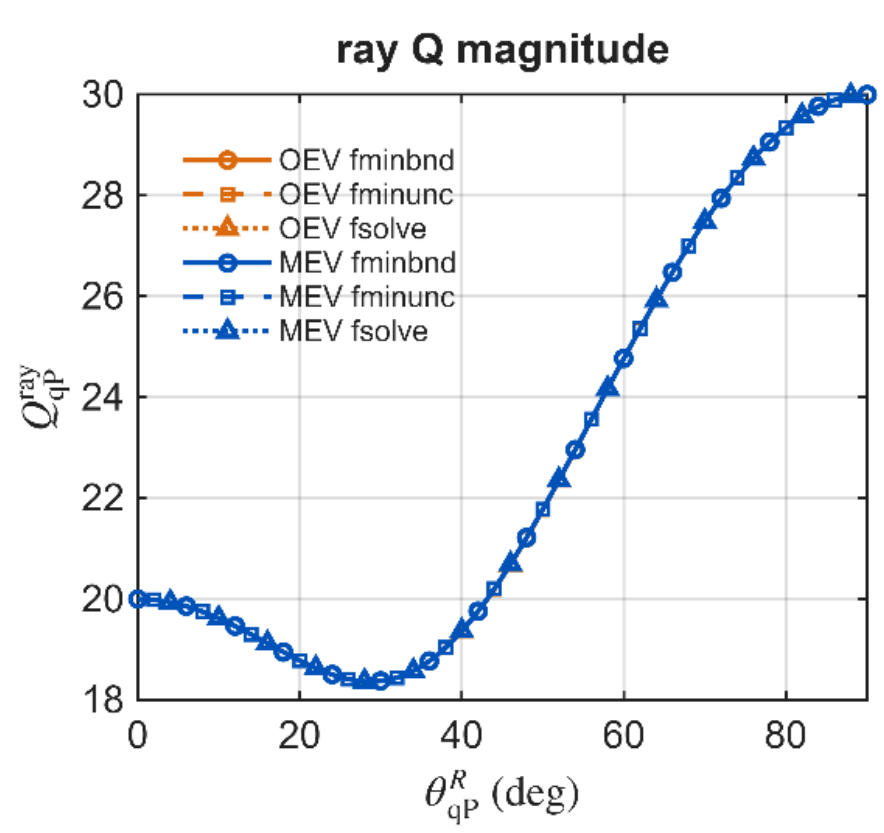
ray Q magnitude
$Q^{ray}_{qP}$
$\theta^R_{qP}$ (deg)
OEV fminbnd
OEV fminunc
OEV fsolve
MEV fminbnd
MEV fminunc
MEV fsolve

a.

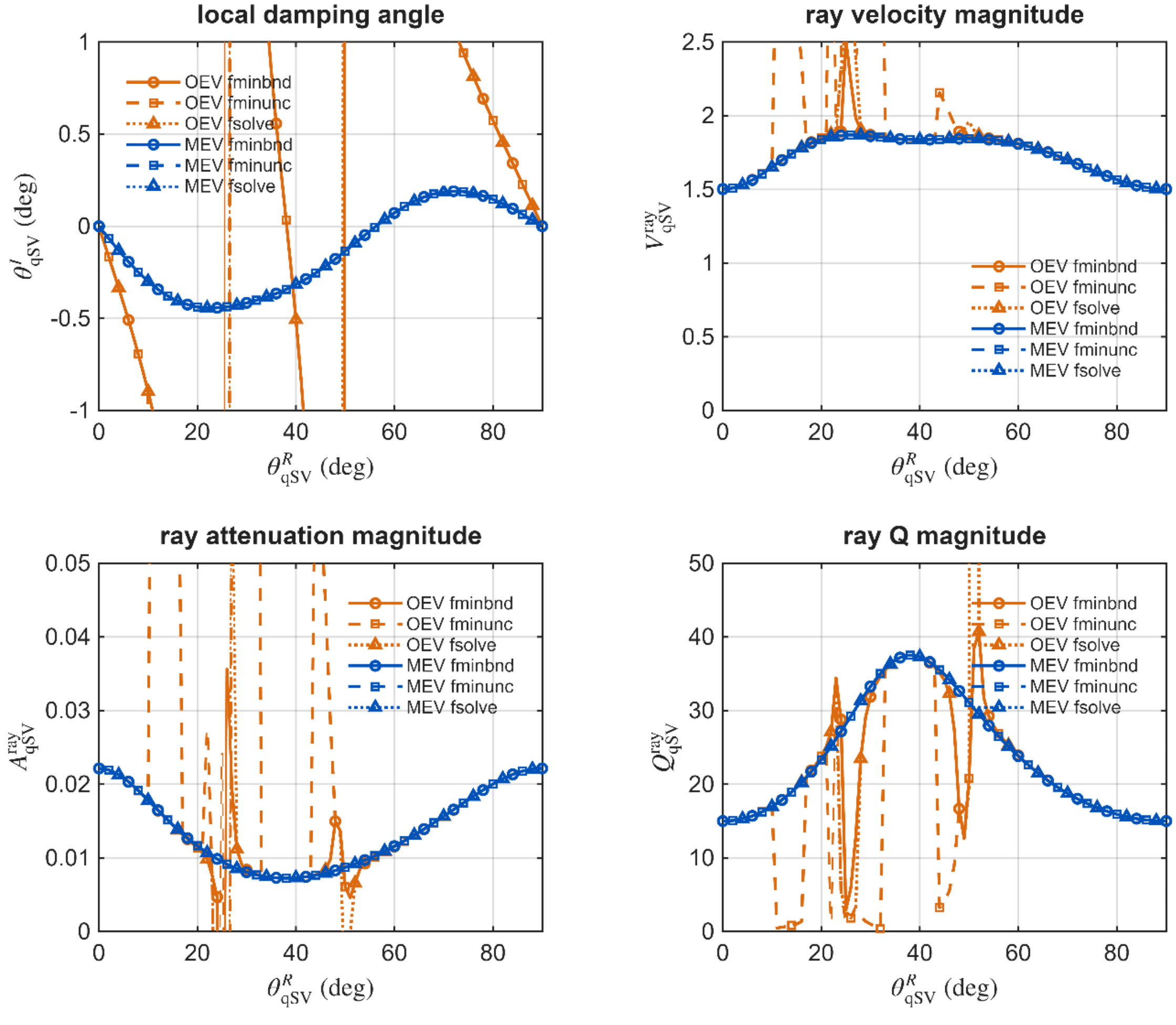

b.

Figure 1. a. qP local benchmark and optimizer selection effect analysis. All of the results from both OEV and MEV are coherent and stable, even though they were obtained from different sets of imaginary slowness direction angles. Therefore, the qP serves as the controlling wave mode that must remain the same regardless of enhancements made at the nodal or global level. b. qSV local benchmark and optimizer selection effect analysis. The OEV yields poor performance, as evidenced by the angle pairs and ray quantities, and is highly sensitive to the choice of optimization function, whereas the numerical fidelity comes from the MEV results, as evidenced by the stable, smooth results.

TABLE 1. Rock properties on 4 layers of a 2D Viscoelastic VTI model. The values on the first layer are also used for a local benchmark experiment

| Layer$^{th}$ | Density-normalized Modulus $a_{pq}^{(R)}\left(\frac{km}{s}\right)^2$ | | | | | Quality Factor $Q_{pq}$ | | | | |
|---|---|---|---|---|---|---|---|---|---|---|
| | $a_{11}$ | $a_{13}$ | $a_{33}$ | $a_{44}$ | $a_{66}$ | $q_{11}$ | $q_{13}$ | $q_{33}$ | $q_{44}$ | $q_{66}$ |
| $1^{st}$ | 14.40 | 4.50 | 9.00 | 2.25 | 3.00 | 30.0 | 15.0 | 20.0 | 15.0 | 12.0 |
| $2^{nd}$ | 21.60 | 6.75 | 13.50 | 3.38 | 4.50 | 45.0 | 22.5 | 30.0 | 22.5 | 18.0 |
| $3^{rd}$ | 28.80 | 9.00 | 18.00 | 4.50 | 6.00 | 60.0 | 30.0 | 40.0 | 30.0 | 24.0 |
| $4^{th}$ | 36.00 | 11.25 | 22.50 | 5.63 | 7.50 | 75.0 | 37.5 | 50.0 | 37.5 | 30.0 |

**Field-scale benchmark**

The next experiment evaluates the numerical performance of CHERT. Accordingly, CHERT is compared with the non-cached baseline method and the modified shortest-path method (MSPM), which has been applied extensively in an anisotropic, smooth, viscoelastic medium (Zhou and Greenhalgh, 2005, 2006; Bai et al., 2007; Li et al., 2020; Wu et al., 2021b). Both methods are evaluated under the same optimization function (fminbnd), the same MEV kernel, and the same 2D layer geometry and acquisition configuration. As a result, any possible differences in the raypath and traveltime field reflect purely from field-scale method performances rather than from the constitutive kernel. In addition to the directional cache retrieval and direct nodal lookup, the current CHERT computation leverages heap-based cell scheduling and prevents duplicate nodal evaluation, while the MSPM scans all uncomputed cells or blocks and evaluates all possible node pairs in order. Four siliciclastic layers are used, with the first layer exhibiting increasing stiffness modulus and quality factor values (Table 2). Strong anisotropic stiffness moduli are used in all layers to emphasize the cusps on the qSV wavefront, while significant energy loss is observed in the first and second layers, as shown by the small Q-Factor values that increase with depth. Surface acquisition, where the source and receivers are on the surface, is used for this test (Figure 4).

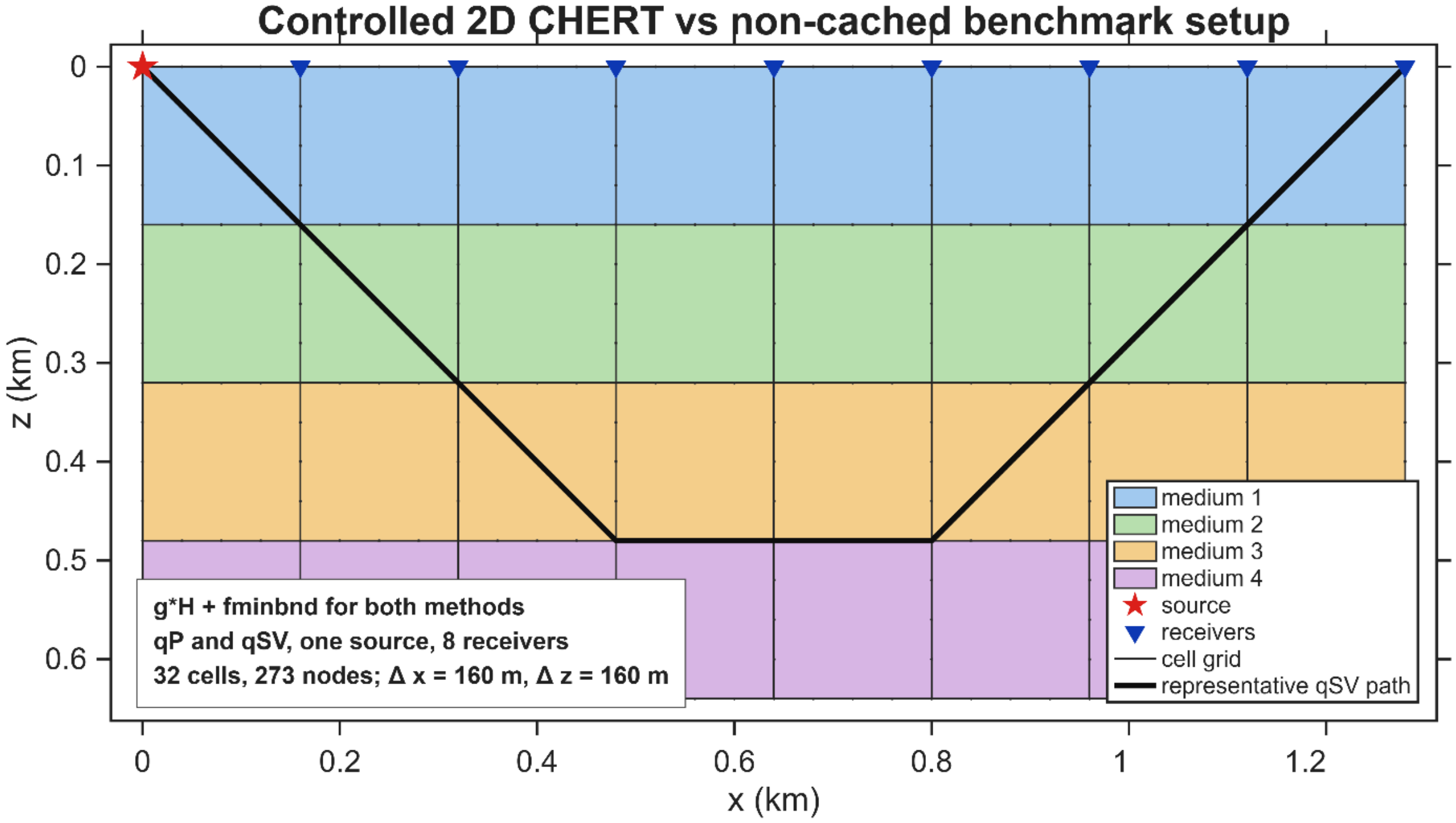

Figure 4. Controlled 2D viscoelastic VTI media for field-scale benchmark. A simplified layered model is used to capture the algorithmic effect on cache usage while avoiding the expensive, potentially runtime-consuming MSPM.

The qP traveltime and raypaths serve as a positive control, where both the propagation and attenuation in the traveltime fields from CHERT and MSPM are qualitatively the same, and their differences are located at the numerical noise level below $10^{-13}$ ms (Figure 5). The overall attenuation traveltime range from both methods is markedly lower (0 to 4 ms) than the propagation traveltime values (up to 300 ms), indicating low energy loss. Both the propagation and attenuation traveltime fields increase with increasing distance from the source, depicting the behavior of the qP wavefront with respect to lateral and vertical variations. Similarly, the propagation and attenuation raypaths of qP from both methods are effectively indistinguishable at every receiver, where direct waves are mostly recorded from all receivers (Figure 5). The refraction events are recorded at the first-layer boundary at distances of 1120 and 1280 m from the receiver. These findings imply that cache utilization does not alter the first-arrival ray-propagation mechanism or its modeling.

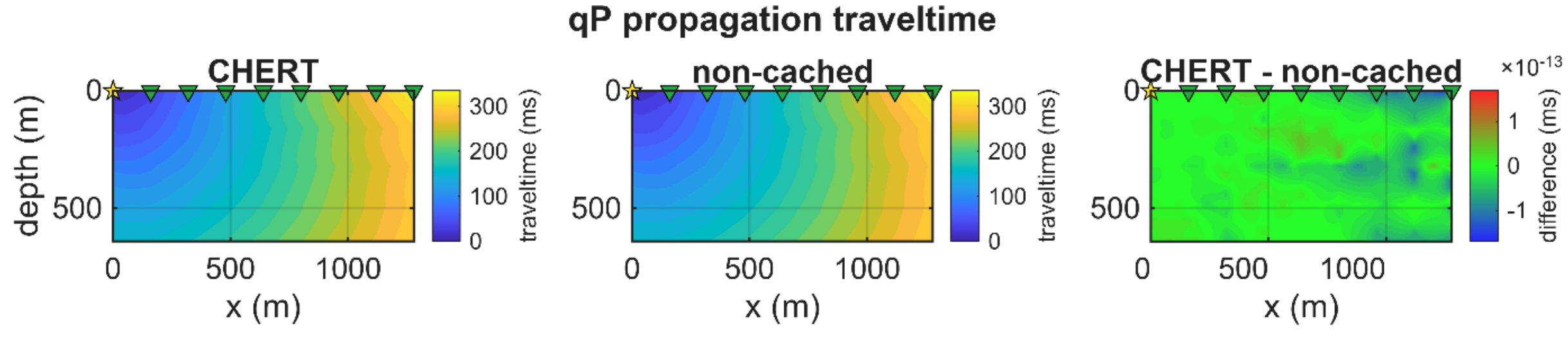

a.

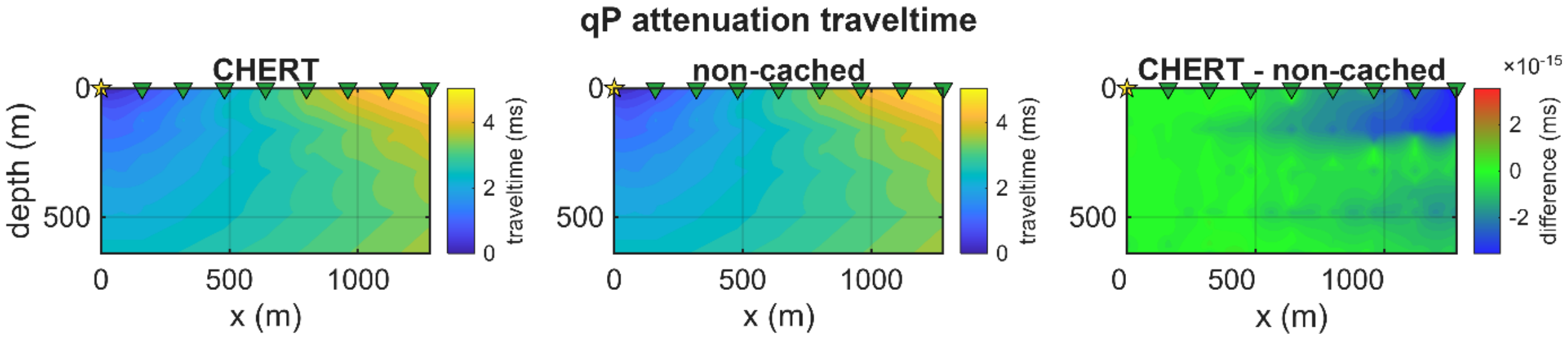

b.

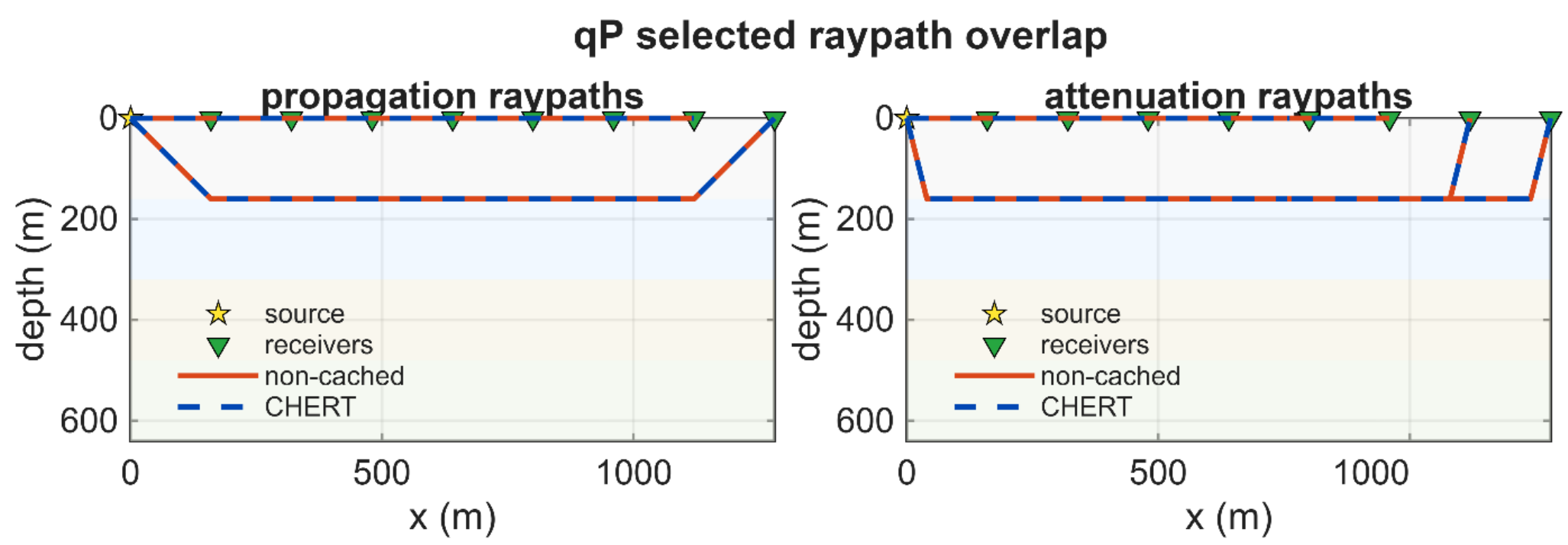

c.

Figure 5. qP field-scale benchmark: a. Propagation and b. Attenuation Traveltime contour, followed by its c. Propagation and Attenuation Raypaths. Note that CHERT and non-cached (MSPM) show visible similarities, with very small differences. Therefore, the qP is still considered the controlling wave mode at the field-scale.

The qSV benchmark provides a more stringent diagnostic test than qP, as small variations in qSV ray quantities, especially in ray attenuation, can significantly alter directional selection (Figure 3). Overall, the propagation traveltimes from both methods have a maximum of 600 ms, consistent with the slower qSV velocity relative to qP. However, their propagation traveltime fields yield 5 ms discrepancies that vary laterally within the same layer and are prominent at 600 m from the source, both laterally and vertically. These discrepancies affect ray trajectories in qSV, as slight changes in traveltime can influence node-to-node trajectory selection. The overall attenuation traveltime yields a greater energy decay than qP, with a maximum of 20 ms. Similar patterns, however, are observed: the smaller, more localized misfit values are present in the attenuation in traveltime field results, with values up to 0.2 ms located only near the surface, away from the source (Figure 6). The qSV propagation raypaths remain broadly consistent for all receivers, with the direct waves traced from the first two receivers, while the refraction events are modeled at the first, second, and third layer boundaries on the remaining receivers with larger offsets. Similarly, the qSV attenuation raypaths are broadly identical at all receivers, with a direct wave traced from the first receiver, while the refraction events are located at the same layer boundaries. (Figure 6).

The runtime indicates that CHERT completed the simulation in around 5 seconds, while the MSPM took almost 329 seconds to complete the same computation. This result demonstrates that CHERT reduces computational cost while maintaining comparable solution quality. Cache-based computation substantially reduces computational cost while preserving the qP raypaths and maintaining comparable qSV solutions. Using the same nodal MEV solvers with MSPM, CHERT effectively preserves indistinguishable raypath and traveltime results in qP and qSV. These findings support the pointwise kernel truth and angular-coverage tests for comprehensive validation.

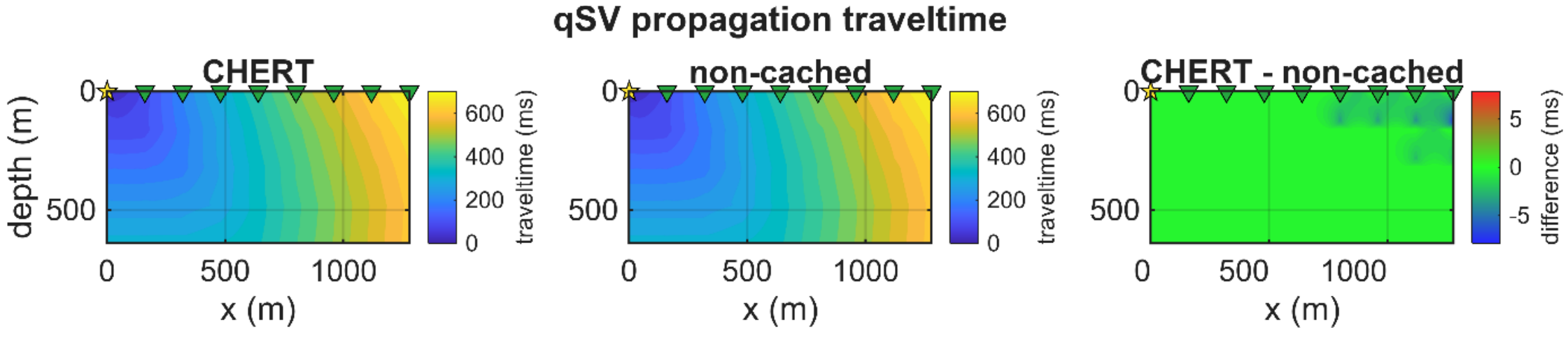

a.

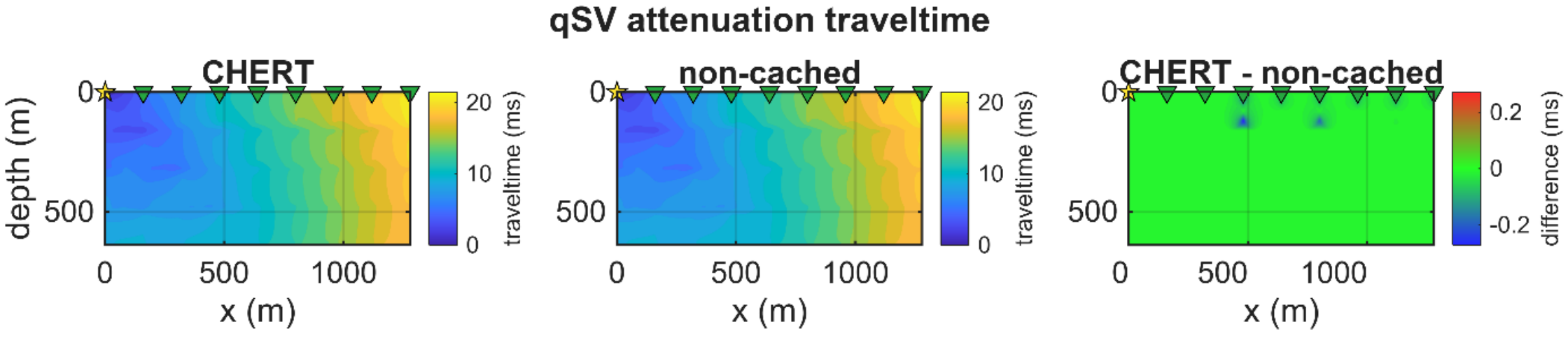

b.

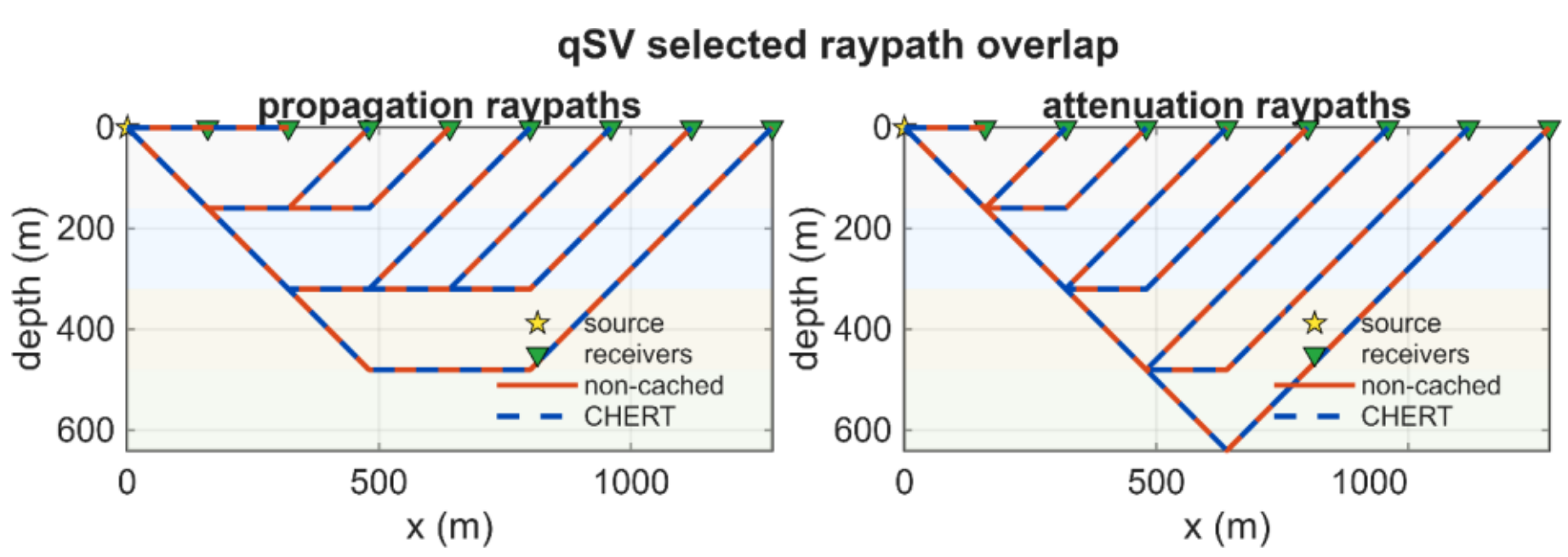

c.

Figure 6. qSV benchmark comparison of a. propagation and b. attenuation with their difference map (right) from CHERT (left) and non-cached baseline or MSPM (middle), followed by c. Simplified surface configuration benchmark of qSV propagation (left) and attenuation (right) in raypaths from CHERT and non-cached (MSPM). The propagation and attenuation discrepancies are qualitatively similar, with small lateral propagation differences, while the attenuation discrepancies are negligible and localized. Both methods remain broadly coherent. The propagation and attenuation raypaths from both methods are generally similar. These findings encourage comprehensive angular-coverage analysis for validation tests.

**Nodewise local-truth and angular-coverage validation**

To ensure the consistent traveltime fields and raypaths of qP and qSV across both methods, two complementary tests are conducted. The first complementary test is a nodewise local-truth benchmark versus a nodal-scale MEV solution as the reference. We utilize $1^0$ directional sampling for this validation test. The second one is an angular coverage benchmark to detect any valid directionality path losses.

The nodewise local-truth benchmark compares CHERT and MSPM against the local MEV reference at representative smooth, interface, and qSV-sensitive nodes. The qP results from CHERT and MSPM are consistent with the nodal solution across the dense VTI angular variation, providing additional evidence that this wave mode serves as a controlling parameter (Figure 7). For qSV comparison, CHERT is consistent with the local MEV reference at these nodes and gives lower RMS errors than MSPM on nodes 36 and 16. CHERT's results are comparable with MSPM, while the results from the other two nodes show that CHERT remains tied with MSPM and MEV (Figure 8).

The angular coverage assessment indicates that both methods can achieve full angular coverage at the selected node locations, with complete coverage ratios, null missing-angle ratios, and null branch mismatches for qP and qSV wave modes. Based on these findings, any discrepancies in qSV raypaths and traveltime from CHERT are better considered as sampled directional ray-quantity manifold and accumulated field-scale decisions while propagating across the same properties.

**Application in modified Marmousi2**

The application in modified Marmousi2 serves as a structurally complex 2D test of the CHERT implementation. The model is cropped to 16 km x 3.5 km by removing 500 m of the upper part, comprised mainly of a water column, discretized with a 30 m x 30 m cell size, with a total of 288,340 nodes and 57,267 cells. Since the original Marmousi2 provides $V_P$, $V_S$, and density (Martin et al., 2006), the current experiment extracts only $V_P$ from dataset. The $V_{qSV}$ and $V_{qSH}$ are modeled using the plausible shear-wave splitting ratio with qP ($V_{qSV} = \frac{V_{qP}}{1.9}, V_{qSH} = \frac{V_{qP}}{1.8}$) on siliciclastic rocks, where qSH is slightly faster than qSV in this case. The missing attenuation parameters are assigned using empirical scaling constants for quality factor and attenuation modeling to keep attenuation magnitudes computationally stable and interpretable for the model demonstration. Since the water column is removed, the source and receiver configurations follow the surface acquisition settings, with 568 receivers employed in the modeling at a 30m spacing.

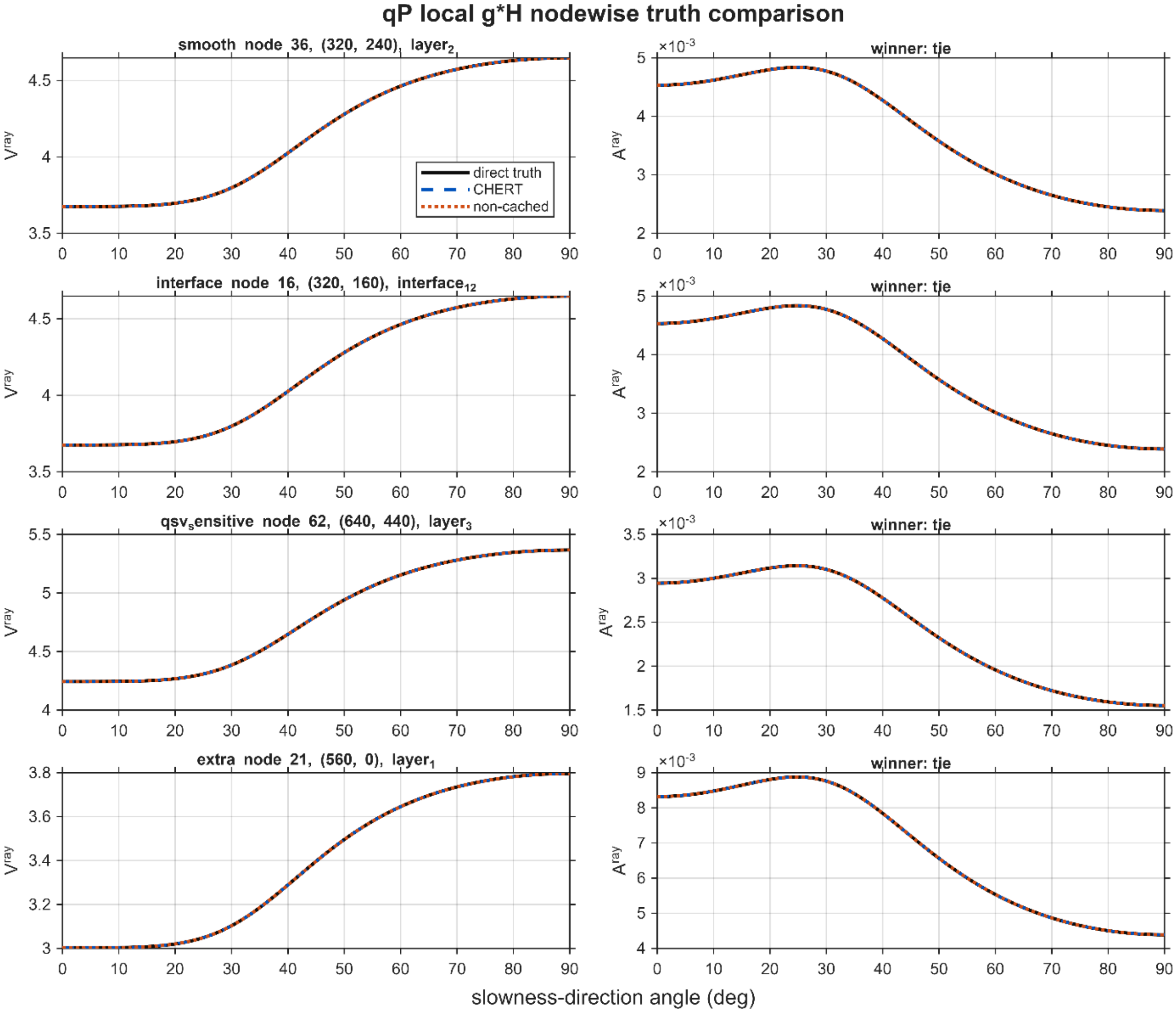

Figure 7. qP kernel-scale $V^{Ray}$ and $A^{Ray}$ comparisons for CHERT and non-cached (MSPM) relative to MEV, on four investigative nodes, which are a smooth internal layer node, a boundary layer node, a qSV-sensitive node, and additional diagnostic nodes, with selected diagnostic-node comparisons. Note that results are largely consistent across all nodes, validating qP as the controlling wave mode.

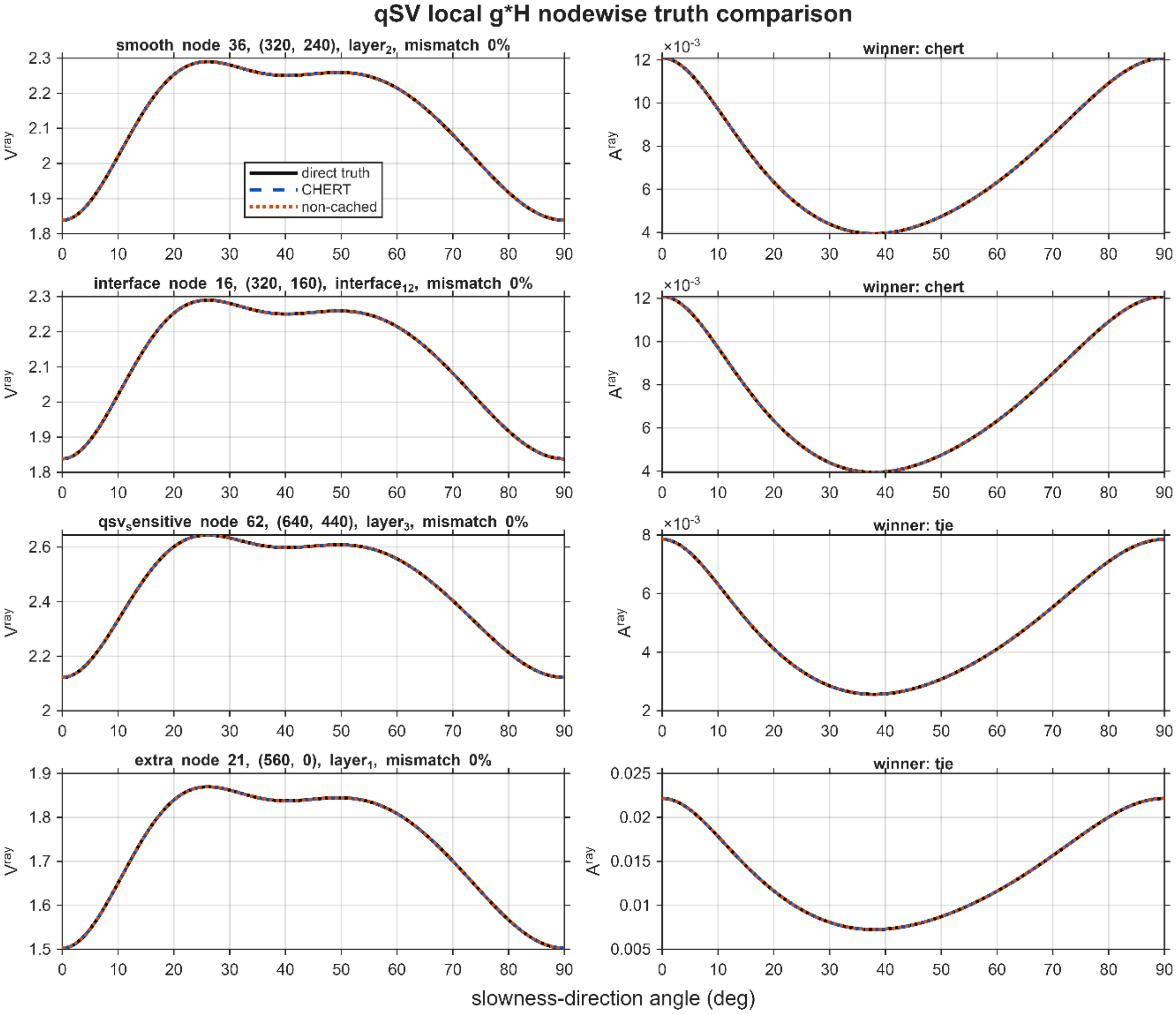

Figure 8. qSV kernel-scale $V^{\text{Ray}}$ and $A^{\text{Ray}}$ comparisons for CHERT and non-cached (MSPM) relative to direct truth (MEV) on the same nodes as in Figure 7. Although similarities also occur, at nodes 36 and 16, CHERT has smaller RMS errors than traditional MSPM. These comparisons support the cache consistency relative to the MEV as the reference model, while the qSV raypath behavior is interpreted separately through the angular-coverage analysis.

The first-arrival traveltime fields remain smooth for qP and qSV propagation and attenuation traveltime (Figure 9). The qP and qSV raypath results, in both propagation and attenuation, show possible first-arrival events, such as diving waves and refractions, that correspond to the model's structural complexity. The raypath coverage and density across all results show a similar pattern on the left side of the source. On the right side of the source, however, the raypaths are substantially denser and propagate deeper at shorter distances from the source across all qP and qSV results. Meanwhile, the qSV attenuation trajectories have lower raypath density and coverage, implying greater sensitivity of path selection to variations in attenuation-related quantities (Figure 10). The total runtime for full-scale Marmousi2 is 530 seconds.

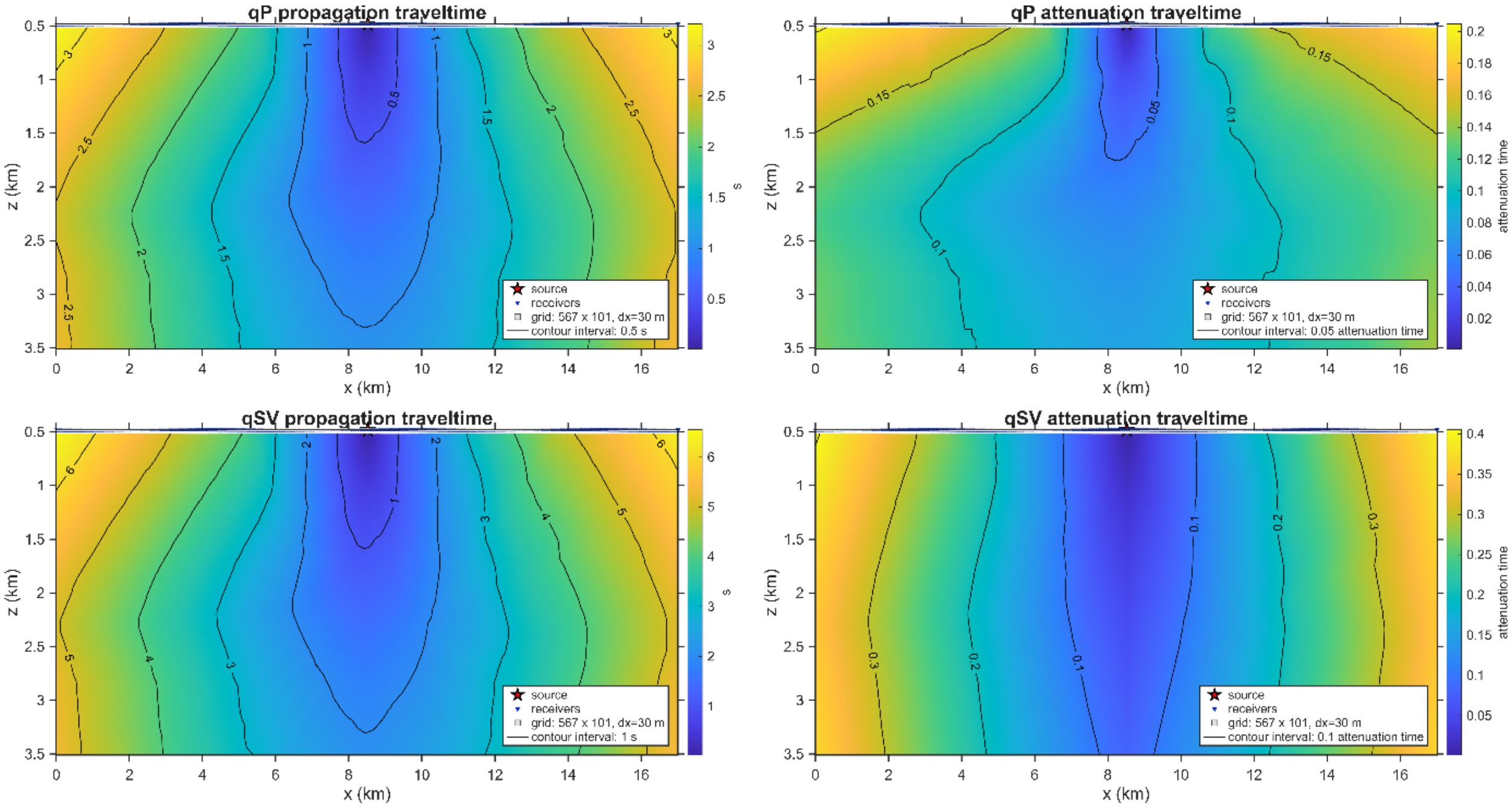

Figure 9. qP and qSV Propagation (left column) and Attenuation (right column) in traveltime generated using CHERT. The qP traveltime field remains smooth and generally coherent, while the qSV is more susceptible to geological complexities. These figures demonstrate the applicability of CHERT to a full-scale synthetic benchmark model.

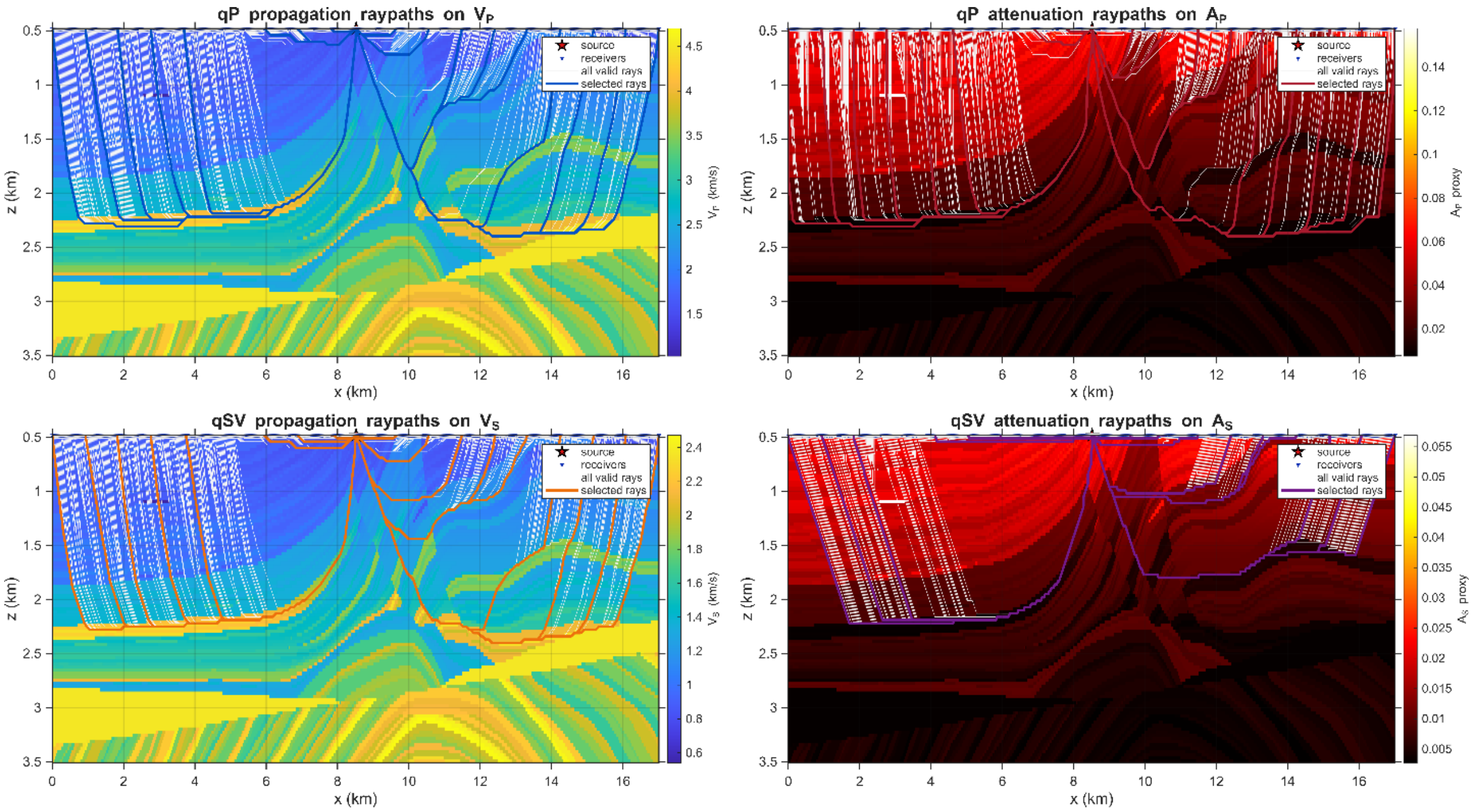

Figure 10. qP and qSV Propagation (left column) and Attenuation (right column) raypaths generated using CHERT overlaid on the related $V_{qP}^{Ray}$, $V_{qSV}^{Ray}$, $A_{qP}^{Ray}$, and $A_{qSV}^{Ray}$ background. Note that the selected rays (thick lines) are highlighted for further analysis. All models exhibit distinct, specified raypath behavior for each wave mode. Note that the attenuation trajectories of qSV have distinct, smaller ray densities compared to others, and miss the thin gas channel with high attenuation on the left part of the Marmousi2.

**Compact 3D TTI extension**

The 3D extension is presented to demonstrate the extension from 2D visco-VTI to 3D visco-TTI. Here, we use the same 4 layers and parameters as in Table 2, with additional tilt angles. The tilt angle is 25 ° across all layers, with undulating layer boundaries. The total model size is 2880 m x 640 m x 2880 m, with a 160 m x 160 m x 160 m cell size, 1296 blocks, and 3024 nodes. The survey acquisition geometry still uses a surface configuration, with a source and 361 receivers.

The computation was performed on qP, qSH, and qSV, but we show only the qSV propagation and attenuation, as the most demanding wave mode in our numerical modeling.

The attenuation and propagation traveltime field shows coherent behavior, with the largest traveltime evenly distributed along the model edges. The attenuation in traveltime units remains small due to the high Q-Factor values, which are inversely proportional to the large initial Q-Factor used as the initial parameter (Figure 11). Meanwhile, the propagation raypath and attenuation in the raypath unit depict the first-arrival events that follow the undulating boundaries of the deepest layer and are traceable from receivers to the source. This example demonstrates that the CHERT application can be extended from 2D VTI to 3D TTI and can also handle irregular layer boundaries.

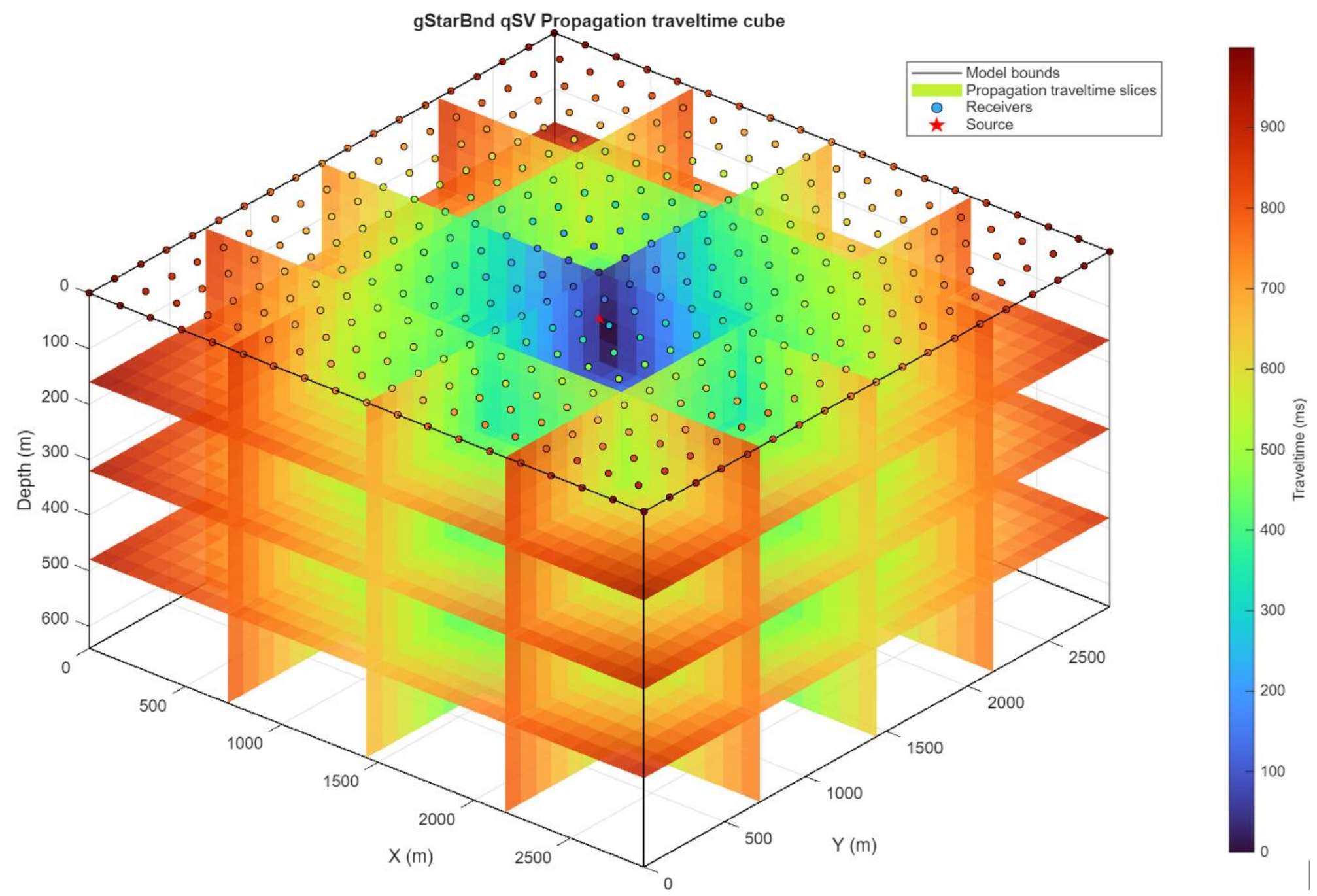

a.

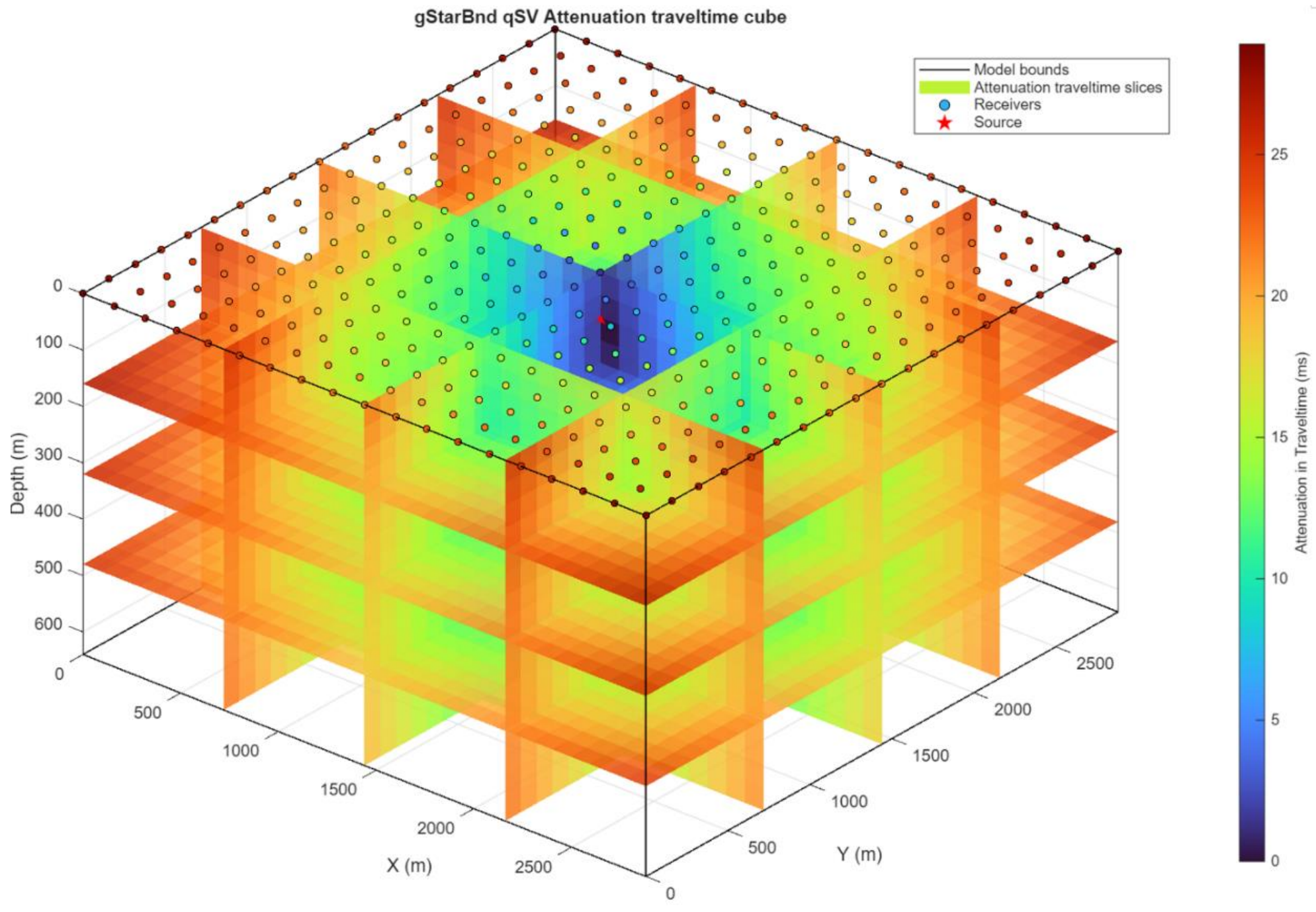
gStarBnd qSV Attenuation traveltime cube
Model bounds
Attenuation traveltime slices
Receivers
Source
Attenuation in Traveltime (ms)
Depth (m)
X (m)
Y (m)

b.

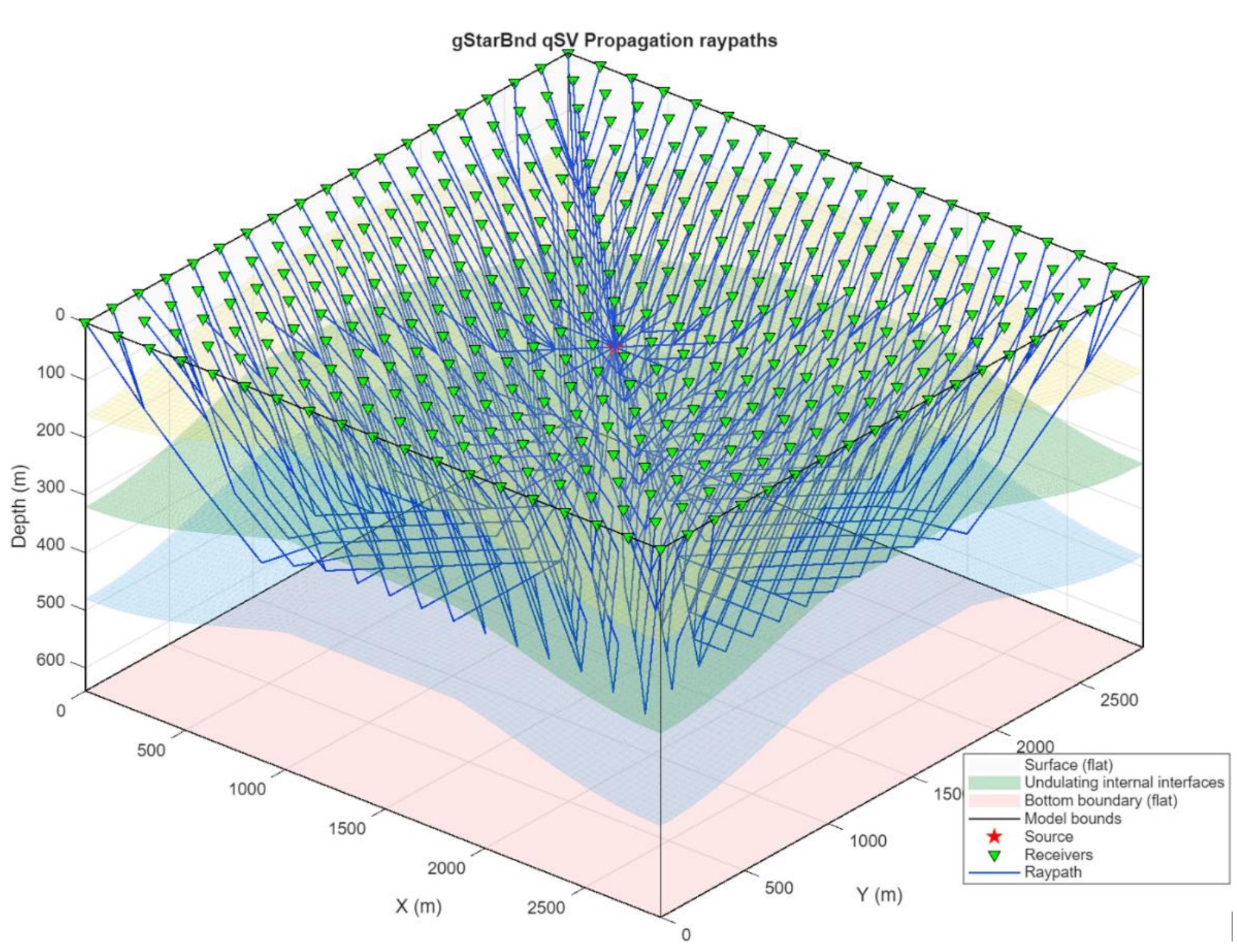
gStarBnd qSV Propagation raypaths
Surface (flat)
Undulating internal interfaces
Bottom boundary (flat)
Model bounds
Source
Receivers
Raypath
Depth (m)
X (m)
Y (m)

c.

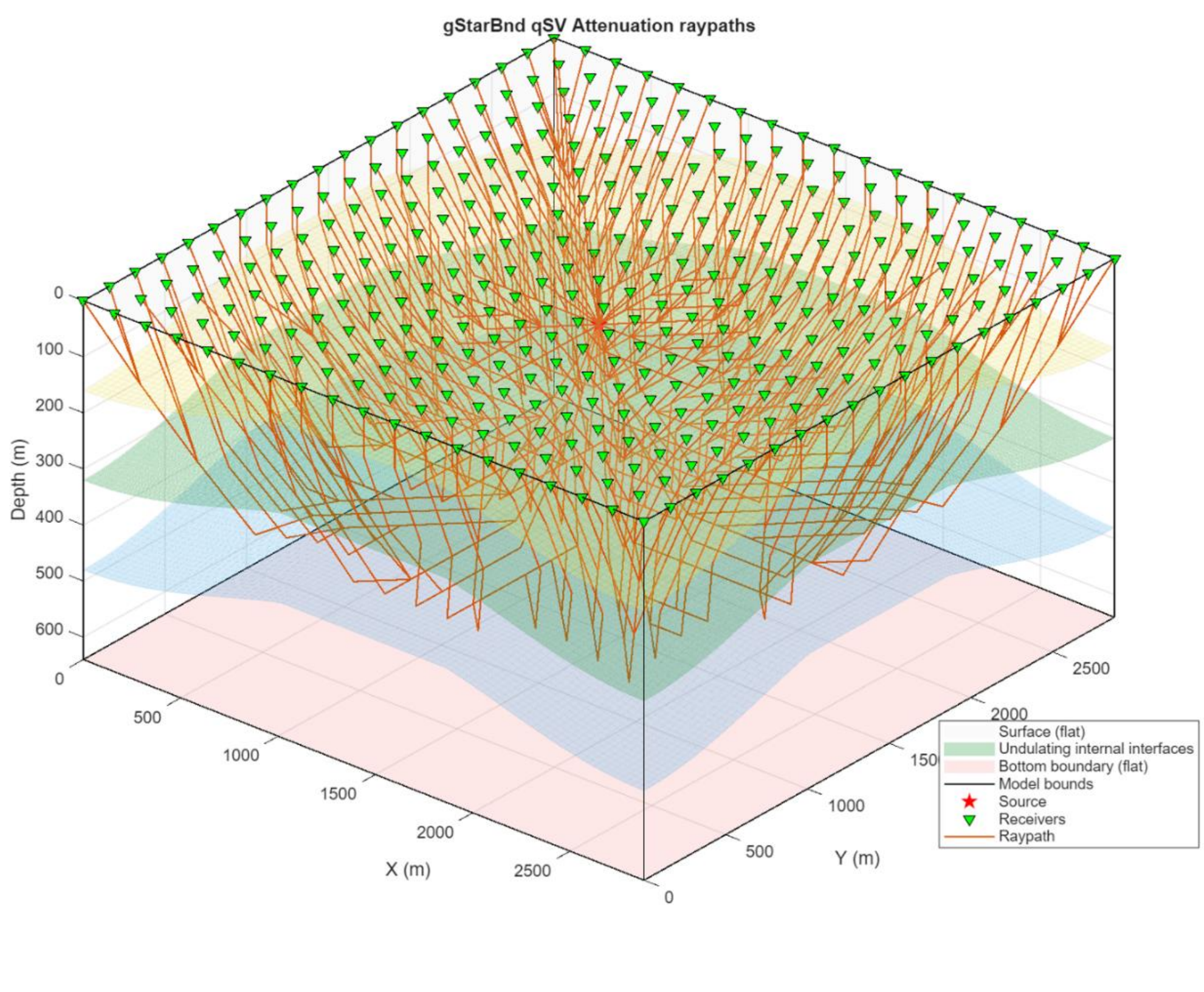

d.

Figure 11. The results of CHERT's application, showing a. propagation and b. attenuation in traveltime in 3D-TTI, driven by MEV, followed by c. propagation and d. attenuation in the raypath in 3D-TTI, driven by MEV. These traveltime and raypath models demonstrate CHERT's natural extensions from 2D to 3D.

**Ablation Studies on CHERT**

The first ablation study aims to examine the core of CHERT's directional discretization. In comparison with the fine-sampling reference, the qP misfit values remain small, indicating the consistent accuracy even when the angular step sampling is enlarged to every $10^0$ (Figure 12). The qSV RMS error values, in contrast, increase more abruptly when the sampling step is coarsened above $2^0$. This finding indicates that qSV results may deteriorate when grid sampling is too coarse.

The refraction events also become inaccurate in modeling the correct critical angles, as coarser angular sampling at $10^0$ leads to larger refraction angles, affecting the ray trajectories (Figure 13). Beyond $2^0$ sampling step, even though the qSV propagation error increases strongly, the runtime actually decreases. Thus, to compromise the increasing runtime at the fine sampling step, the $1^0$ sampling step is the suggested practical balance for preserving the qSV accuracy at the cost of half of the $0.5^0$ sampling step (Figure 14).

In our previous Marmousi2 modeling, no facies generalization or grouping was implemented. The parameterization is conducted cell-wise by assigning a specified cache class and the corresponding ray-quantity tabulation to each cell. In practice, efficient computation runtime on large models requires averaging several rock properties into a single rock type or facies. This requirement motivates the second ablation study, which examines the effect of facies or rock-type classifications on the modified Marmousi2 model. Grouping the facies into 12 or fewer groups yields larger qSV discrepancies than cellwise grouping. Averaging into a 24-group case still results in a large misfit compared to the cell-wise case (Figure 15). When grouped into 4 or 12 groups for computational runtime, the qSV propagation and attenuation raypaths produce noticeably different raypaths, with shallow raypath coverage, irregular first-arrival events outside layer boundaries, and lower ray densities than in the cell-wise case (Figure 16). These results demonstrate that the cache class numbering must be defined thoroughly in geologically complex models.

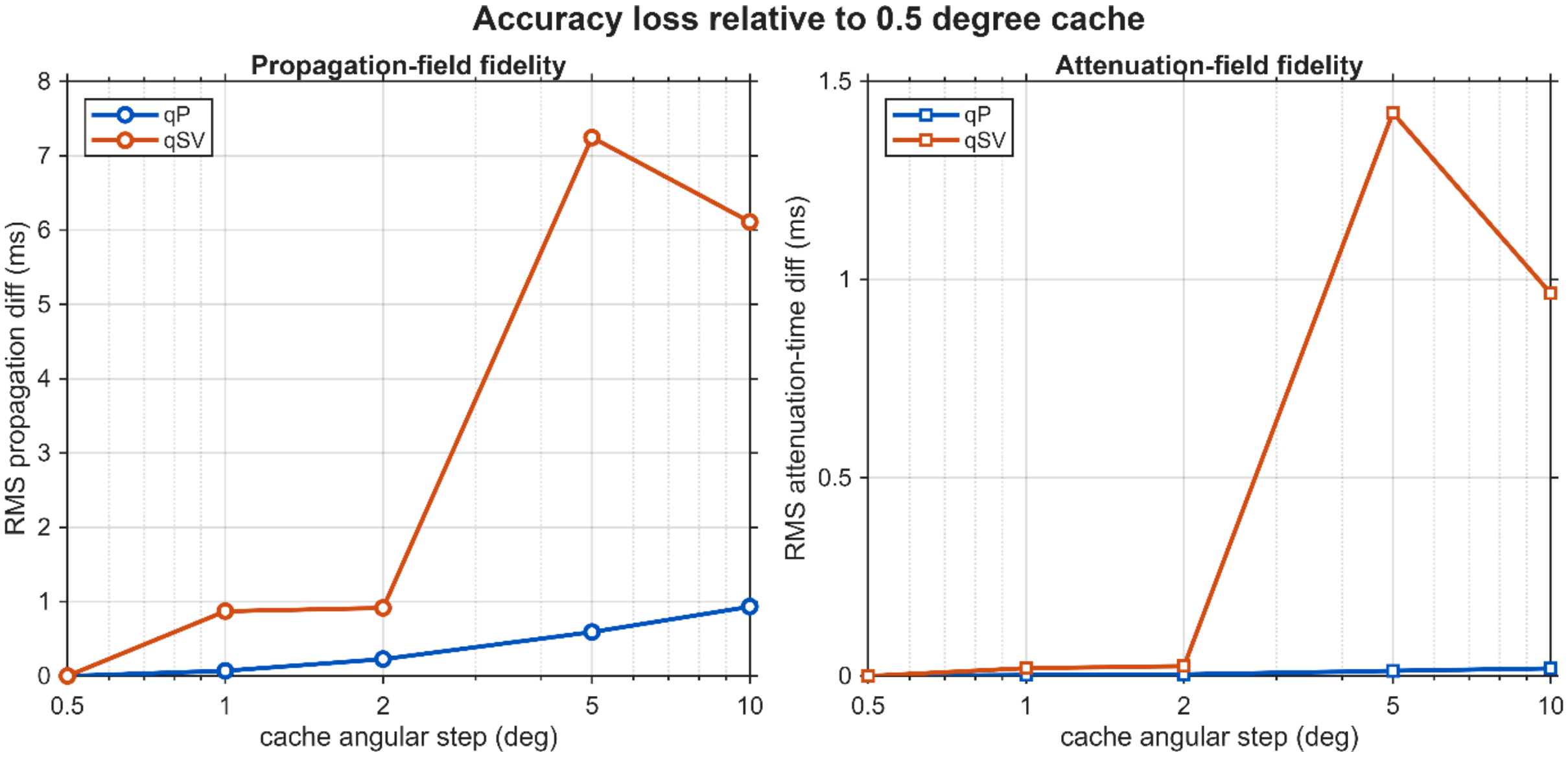

Figure 12. The ablation study on the effect of coarsening the angular step during field-scale modeling using CHERT on the RMS loss of qP and qSV propagation and on attenuation in traveltime. qP error in attenuation field remains very stable despite coarsening angular step, while the error in propagation field shows a subtle increase after $2^0$ coarsening step. Meanwhile, both qSV fields show a major increase after $2^0$ coarsening step, showcasing the qSV as a pivotal indicator for selecting a practical angular step.

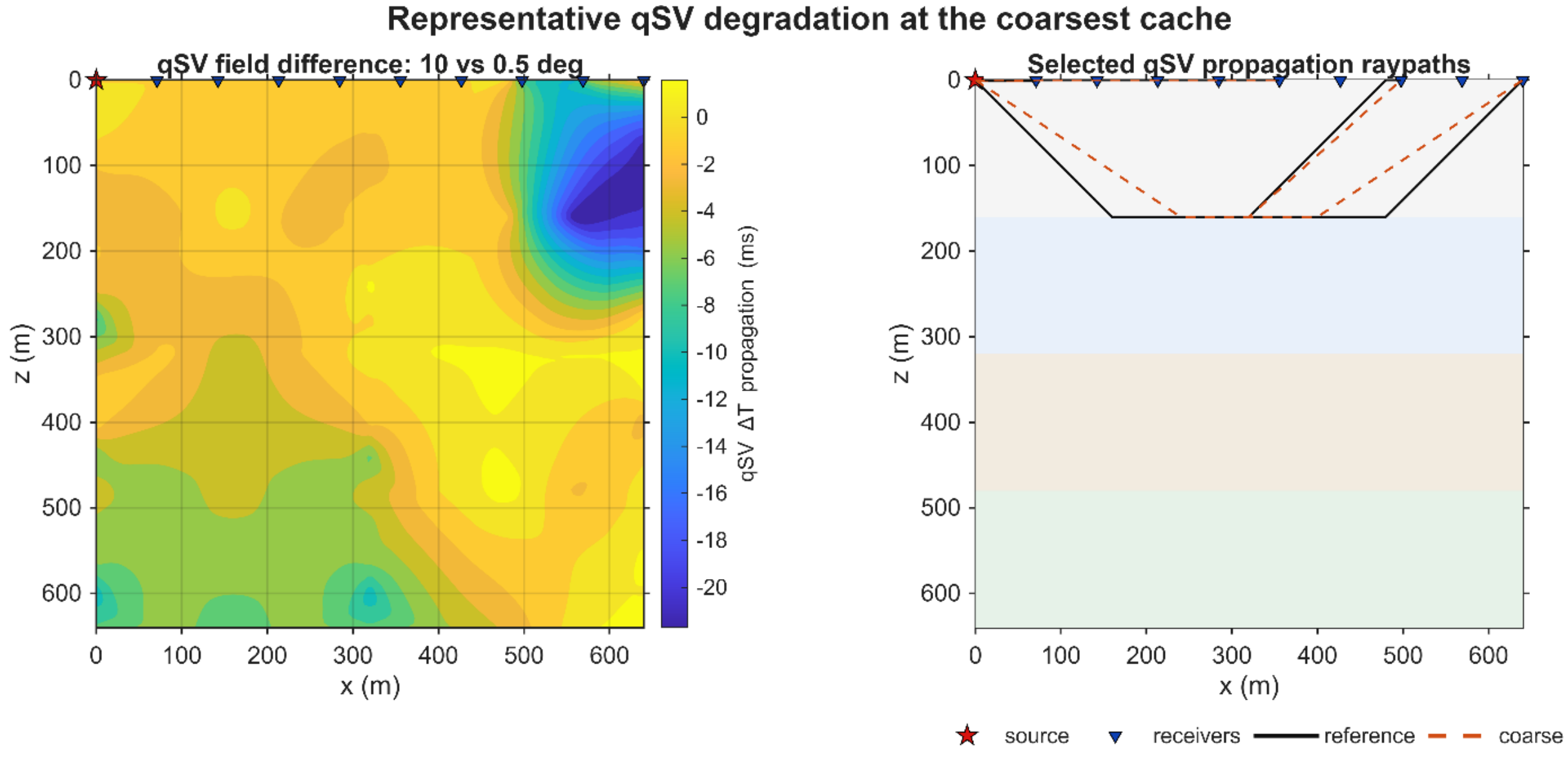

Figure 13. Ablation study of the effect of coarsening the angular step during field-scale modeling with CHERT on qSV propagation traveltime (left) and raypaths (right). The raypath model shows the 0.5° reference-step paths (solid black line) compared with the 10° coarse-step paths (dashed line). Both figures show, physically, how coarsening distorts the traveltime field and the raypath trajectory.

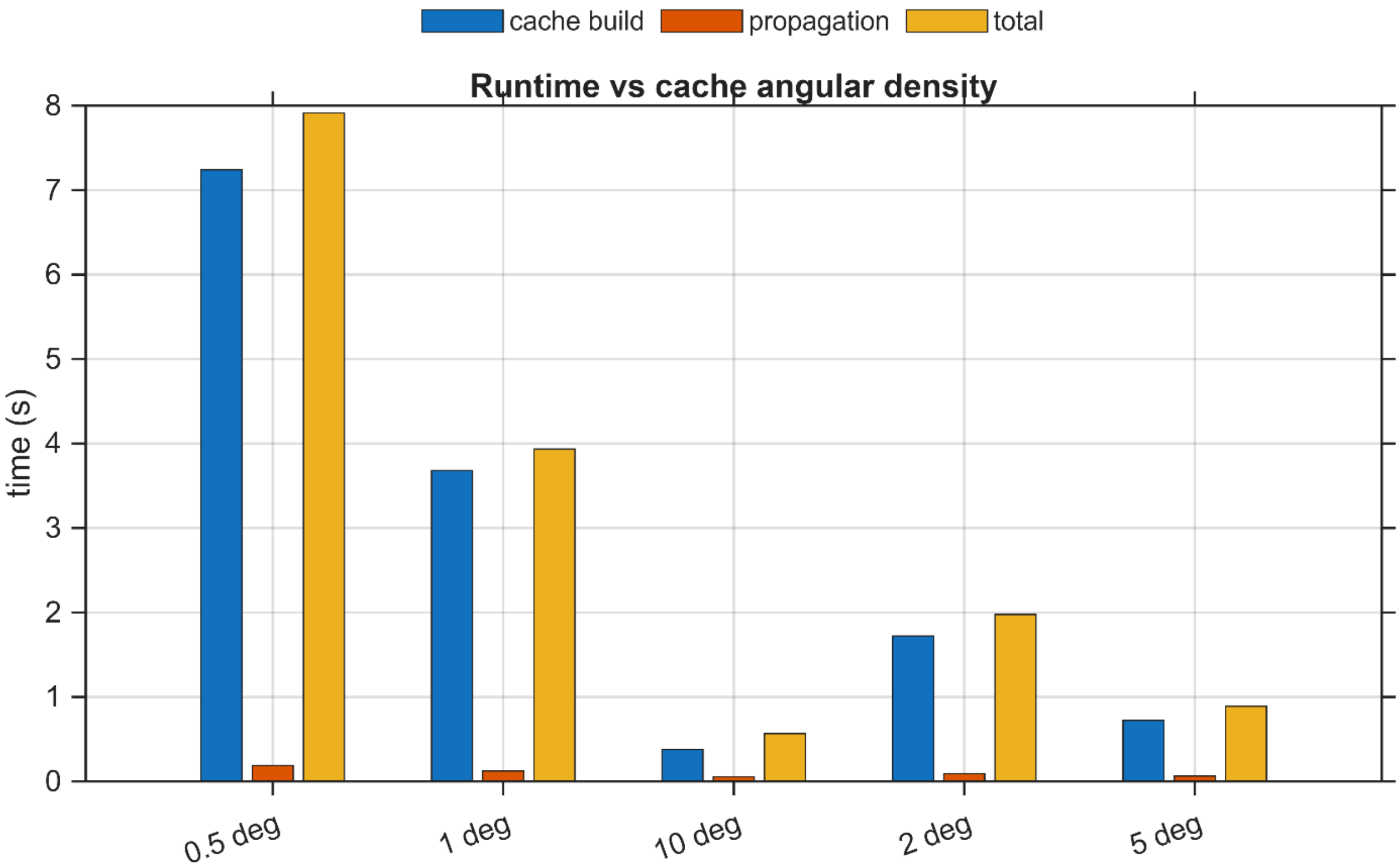

Figure 14. Runtime on various angular step scenarios shown in cache-build time (blue), propagation time (red), and total runtime (orange). Combined with Figures 12 and 13, this figure shows the trade-off between accuracy and cost and supports the final 1° practical configuration.

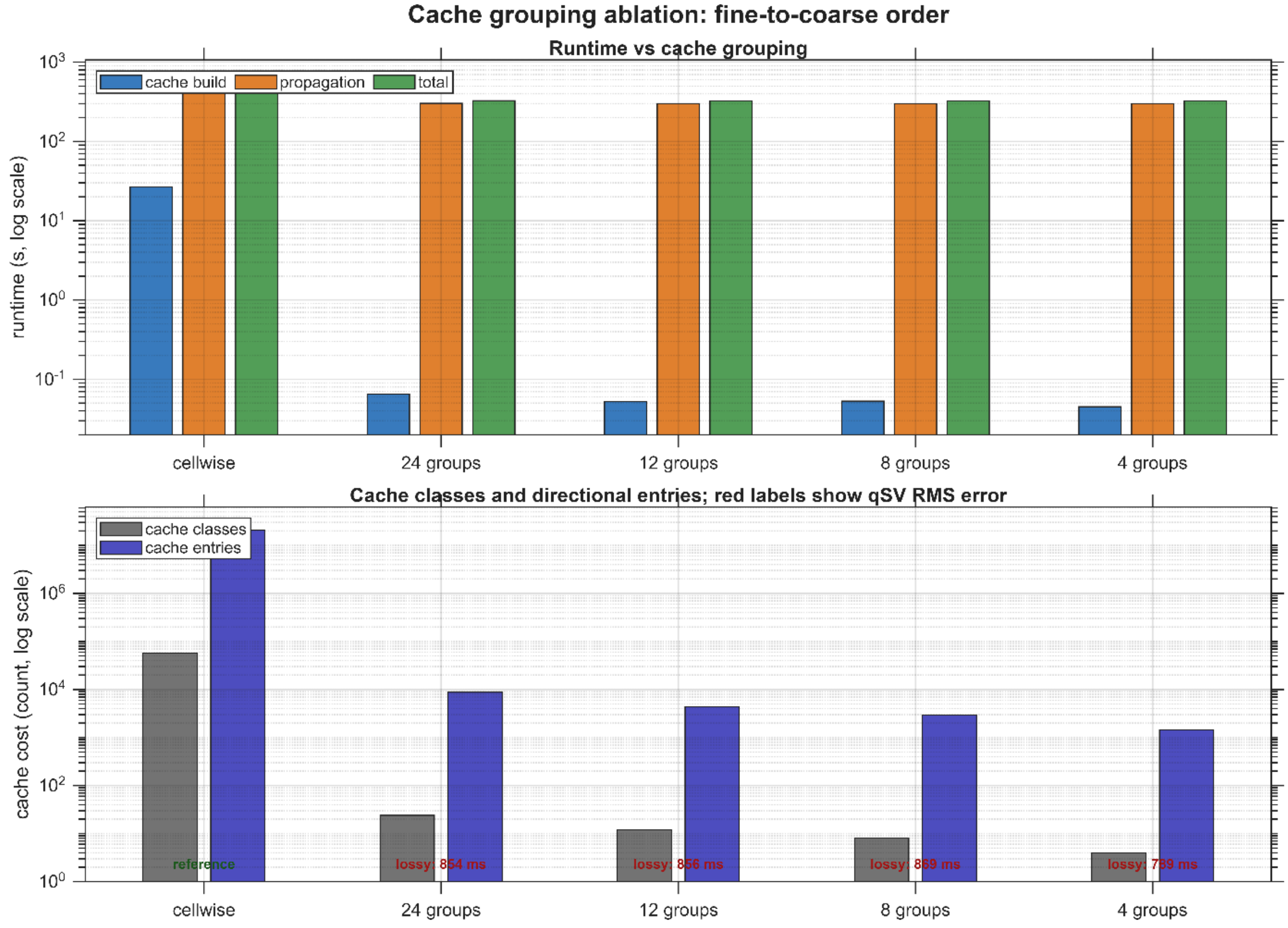

Figure 15. The runtime (upper) and cost summary (lower) comparison across various facies or rock types in the large Marmousi2 model, ranging from cellwise to the coarsest grouping of only 4 rock types. These figures illustrate the trade-off between runtime and accuracy (RMS error) in facies grouping, showing that coarse grouping reduces cache size and build cost but incurs a substantial difference.

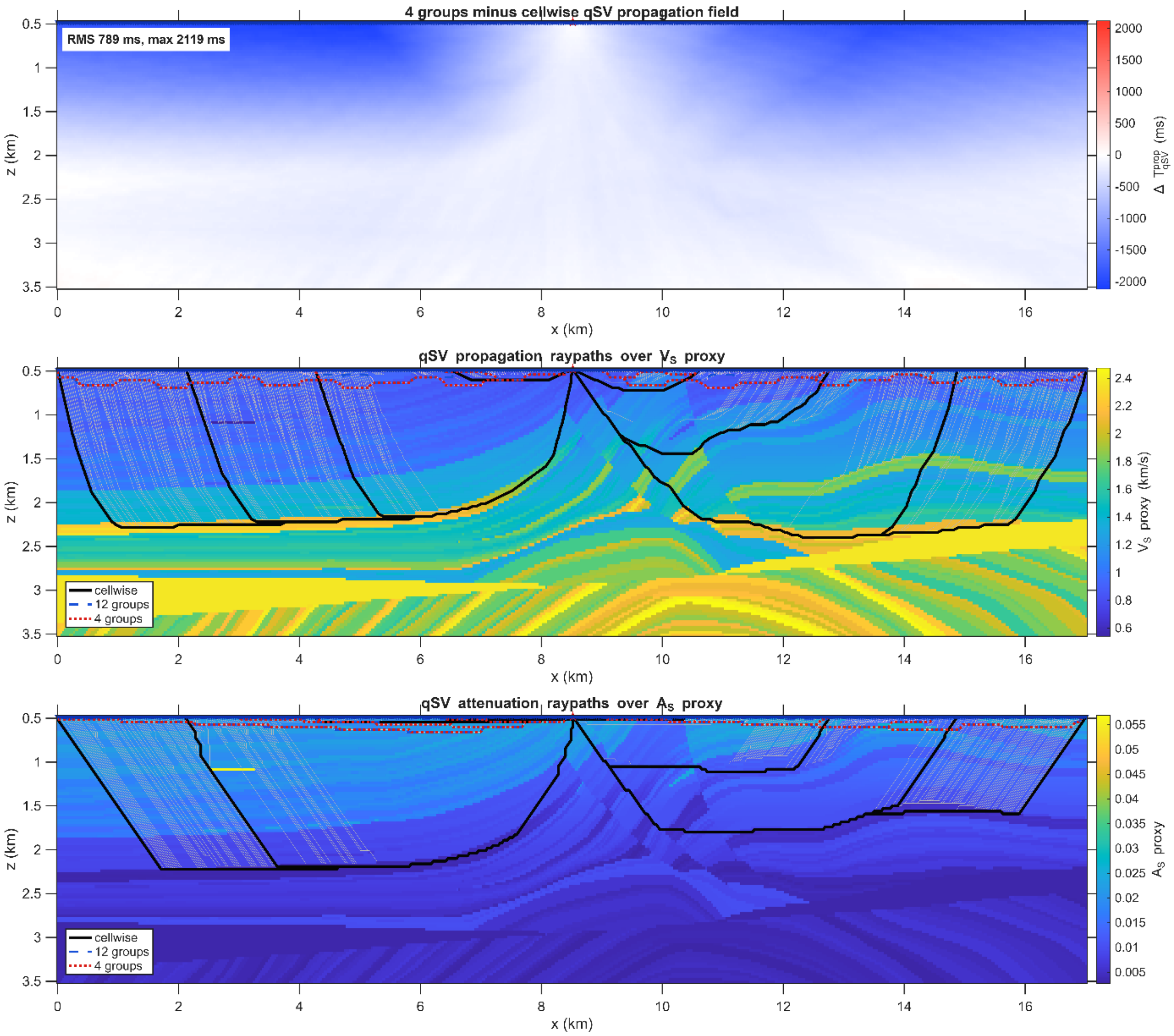

Figure 16. Cache-grouping ablation: representative qSV degradation under coarse grouping. The top panel displays the qSV propagation traveltime field misfit between the 4-group case and the cellwise reference. The 12 and 4 group of qSV raypaths (dashed lines) are compared to the cellwise raypaths (solid lines) in qSV propagation (middle) and attenuation in raypaths modeling. The figure demonstrates that overly coarse spatial grouping alters both qSV traveltime fields and raypath trajectories heavily near the surface (<1km).

# DISCUSSION

The field-scale benchmark shows two distinct behaviors. In qP, CHERT and MSPM are effectively indistinguishable at the plotted results, confirming that the cache does not alter the first-arrival mechanism in the shortest-path logic for this control mode. Meanwhile, the challenging qSV modeling highlights the discrepancy between CHERT and MSPM. In this wave mode, shown in Figure 3, the attenuation ray in qSV shows a subtle variation in the order of $10^{-2}$ with the ray direction angle, indicating path-selection sensitivity for the attenuation ray. Any misrepresented angular coverage in the attenuation values leads to abrupt changes in the field-scale path-search process. In this wave mode, branch sensitivity, the presence of cusps, and the attenuation non-linearity materially govern the raypath trajectory and traveltime fields. The runtime comparison between MSPM and CHERT demonstrates that the proposed method considerably reduces the runtime from 329 s to approximately 5 s.

These observations are further supported by the nodal benchmark test with MEV, which shows that the proposed technique yields a lower RMS error relative to the MEV reference than the MSPM baseline. These findings indicate that CHERT produces consistent nodal-scale behaviors with the MEV reference solution. The angular coverage check also implies that our method does not miss any plausible local directions at the tested node.

The ablation studies show practical CHERT's configurations. Node density controls nodal angular coverage and determines the reliability of the directional manifold representation. Cache-class granularity determines the medium representativeness. The qP remains comparatively tolerant to moderate cache group coarsening, but qSV is more sensitive to both angular sampling and cache-class grouping. This finding is consistent with the prior controlled benchmark, in which qSV is also shown to be the most sensitive mode at both the nodal and field scales.

The directional-density ablation implies a practical design of CHERT. If the angular sampling becomes too sparse, the cached qSV manifold can be underresolved for directional quantities, and the field-scale qSV traveltime field and raypath begin to distort. The grouping-granularity serves a complementary configuration. If physically distinct blocks or cells are modeled with extremely coarse cache classes, the cache can no longer provide an accurate medium representation, and the qSV fidelity of traveltime and raypath again worsens. These results highlight the practical configuration of CHERT's framework and emphasize the importance of careful selection of these configurations to keep the cache representative. For practical implementations, both cache angular sampling and grouping granularity should then be identified from qSV-focused ablation studies, since qSV is the most sensitive mode to both angular undersampling and coarse grouping.

The scope of the present work is intentionally limited. CHERT is formulated and examined for first-arrival applications in viscoelastic VTI and TTI media, and its validation is primarily based on synthetic benchmark models. The 3D and TTI extensions are included as practical evidence of concept rather than as another benchmarking test. The node-level truth comparison and angular coverage analysis are performed on selected nodes rather than the entire medium. Another limitation is that the present numerical observation does not present all existing energy-velocity or ray-quantity solvers as kernels, nor does it present all possible ray-tracing formulations—the local benchmark aims to select a reliable kernel for practical applications of CHERT.

The observed runtime reduction suggests that CHERT may be suitable as an iterative forward solver in seismic tomography, inversion, or seismic migration workflows. The cache architecture may minimize the cost for incorporating viscoelastic anisotropic ray quantities into these applications, as the repetitive forward solver often accounts for the majority of the runtime cost in inverse problems. Thus, the directional-cache strategy provides a potential bridge between physically reliable, computationally efficient ray tracing and viscoelastic, anisotropic inverse problems.

Future studies should aim in two directions. The first is a comprehensive validation of CHERT implementation, including more complex models and field data implementation in 2D and 3D across various configurations and terrestrial settings. The second is the workflow integration of the forward modeling method into a larger seismic inverse problem framework that can examine the application of propagation and attenuation inverse models.

## CONCLUSIONS

This work introduces CHERT, a cached-Hamiltonian enhanced ray tracing method for viscoelastic anisotropic media. Its main contribution is the separation of ray-quantity computation from the shortest-path algorithm through retrievable directional caches. This framework facilitates precomputation of ray velocities, attenuation, quality factors, and directions, tailored to the cache class and wave mode, for subsequent retrieval when updating candidate paths.

The numerical experiments demonstrate that this separation reduces the computation of the Hamiltonian-derived kernel by avoiding repeated kernel calls while preserving the sampled directional ray quantities. In the controlled field-scale benchmark with MSPM, CHERT consistently produces qP and qSV raypaths and traveltimes with the MSPM baseline while reducing computational time from nearly 329 s to around 5 s under the same MEV kernel, model, and acquisition geometry. The nodal benchmark with MEV solutions and angular-coverage tests supports CHERT's nodal consistency at the selected nodes.

The numerical implementations on modified Marmousi2 and 3D TTI indicate the potential to extend and apply CHERT beyond simple 2D layered models, supported by ablation studies that identify practical methodological constraints. In the observed configuration, fine node spacing must be carefully selected to preserve qSV behavior relative to the fine-sampling reference, whereas extremely coarse cache-class classifications can misrepresent the medium's variability. These results imply that, with adequate step-angle coverage and cache-class granularity, CHERT is a practical ray tracing and forward modeling solver for attenuation-aware tomography, migration, and the larger viscoelastic anisotropic inversion workflows.

## DATA AND MATERIALS AVAILABILITY

Data associated with this research are available and can be obtained by contacting the corresponding author.

## APPENDIX A

**Table A-1.** MSPM and CHERT comparison. The main difference lies in how ray quantities are computed: either during node-to-node updates (MSPM) or precomputed (CHERT). The MSPM input is the number of sources $s$ and receivers $r$, the complex density-modulus $a_{ijkl}$, and a discretized grid network $\mathcal{G}$, while the CHERT needs additional cache classes $k$

**Algorithm 1: Conventional MSPM**

Input: $a_{ijkl}, s, r$, and $\mathcal{G}$.
Output: $T_b^{(s,m,R)}, T_b^{(s,m,I)}, \mathcal{R}_{s,r}^{(s,m,p)}$, and $\mathcal{R}_{s,r}^{(s,m,a)}$.
for $s \leftarrow 1, \dots, N$
  for $m \leftarrow 1, \dots, N$ do
    Initiate traveltime and nodes at the source cell
    while uncomputed cells $\neq 0$ do
      scan computed cell and its neighboring cells
      for $i \leftarrow 1, \dots, N\ j\ \&\ \ j \leftarrow 1, \dots, N$ do
        if $i \leftarrow corner, j \leftarrow corner$
          compute $v, V^{Ray}, A^{Ray}, \hat{r}$
        else
          reconstruct $v, V^{Ray}, A^{Ray}, \hat{r}$
        end if
        update $\tau_j^{(s,m,\odot)}, \tau_i^{(s,m,\odot)}$.
      end for
    end while
    trace back the predecessor nodes from $r$ to $s$.
  end for
end for

**Algorithm 2: CHERT**

Input: $a_{ijkl}, s, r, \mathcal{G}$, and $k$.
Output: $T_b^{(s,m,R)}, T_b^{(s,m,I)}, \mathcal{R}_{s,r}^{(s,m,p)}$, and $\mathcal{R}_{s,r}^{(s,m,a)}$.
Cache construction
for $c \leftarrow 1, \dots, N$
  for $m \leftarrow 1, \dots, N$ do
    compute $v, V^{Ray}, A^{Ray}, \hat{r}$
    construct $\mathcal{C}(m, k, \theta)$
  end for
end for
for $s \leftarrow 1, \dots, N$
  for $m \leftarrow 1, \dots, N$ do
    Initiate traveltime and nodes at the source cell
    while uncomputed cells $\neq 0$ do
      scan computed cell and its neighboring cells
      for $i \leftarrow 1, \dots, N\ j\ \&\ \ j \leftarrow 1, \dots, N$ do
        querry $\mathcal{C}(m, k_i, \theta)$ and $\mathcal{C}(m, k_j, \theta)$
        update $\tau_j^{(s,m,\odot)}, \tau_i^{(s,m,\odot)}$.
      end for
    end while
    trace back the predecessor nodes from $r$ to $s$.
  end for
end for